\documentclass[numberedappendix,iop]{emulateapj}
\usepackage[varg]{txfonts}
\newcommand{\se}[1]{\mbox{\S\ \ref{sec:#1}}}
\newcommand{\ses}[2]{\S\ \ref{sec:#1} and \ref{sec:#2}}

\newcommand{\Se}[1]{\mbox{Section\ \ref{sec:#1}}}
\newcommand{\Ses}[2]{\mbox{Sections\ \ref{sec:#1} and \ref{sec:#2}}}
\newcommand{\eq}[1]{equation\ (\ref{eq:#1})}
\newcommand{\eqp}[1]{\mbox{eq.\ [\ref{eq:#1}]}}
\newcommand{\eqs}[2]{equations\ (\ref{eq:#1}) and (\ref{eq:#2})}
\newcommand{\eqsto}[2]{equations\ (\ref{eq:#1})--(\ref{eq:#2})}
\newcommand{\eqps}[2]{eqs.\ [\ref{eq:#1}] and [\ref{eq:#2}]}

\newcommand{\Eq}[1]{Equation\ (\ref{eq:#1})}
\newcommand{\Eqs}[2]{Equations\ (\ref{eq:#1}) and (\ref{eq:#2})}
\newcommand{\Eqsto}[2]{Equations\ (\ref{eq:#1})--(\ref{eq:#2})}
\newcommand{\Eqss}[3]{Equations\ (\ref{eq:#1}), (\ref{eq:#2}), and (\ref{eq:#3})}
\newcommand{\fg}[1]{\mbox{Fig.\ \ref{fig:#1}}}

\newcommand{\Fg}[1]{\mbox{Figure\ \ref{fig:#1}}}
\newcommand{\Tb}[1]{\mbox{Table\ \ref{tab:#1}}}
\newcommand{\app}[1]{\mbox{Appendix\ \ref{app:#1}}}
\newcommand{\vs}{vs.}
\newcommand{\ie}{i.e.,}
\newcommand{\eg}{e.g.,}

\newcommand{\sumi}{(i)}
\newcommand{\sumii}{(ii)}
\newcommand{\sumiii}{(iii)}
\newcommand{\sumiv}{(iv)}

\usepackage{color}
\newcommand{\sub}[2]{\ensuremath{_{\mathrm{#1},#2}}}
\newcommand\labelH[1]{\label{#1}\texttt{\small [#1]}}
\renewcommand{\labelH}{\label}

\shorttitle{Understanding how planets become massive}
\shortauthors{C.W.~Ormel \& H.~Kobayashi}

\begin{document}
\title{Understanding how planets become massive:\\ I. Description and validation of a new toy model}
\author{C. W. Ormel}
\affil{Max-Planck-Institute for Astronomy, K\"onigstuhl 17, 69117 Heidelberg, Germany}
\affil{Astronomy Department, University of California, Berkeley, CA 94720}
\email{ormel@astro.berkeley.edu}

\and
\author{H. Kobayashi}
\affil{Astrophysical Institute and University Observatory, Friedrich Schiller University, Schillergaesschen 2-3, 07745 Jena, Germany}
\affil{Department of Physics, Nagoya University, Nagoya, Aichi 464-8602, Japan}
\email{hkobayas@nagoya-u.ac.jp}
\begin{abstract}
The formation of giant planets requires accumulation of $\sim$10 Earth mass in solids; but how do protoplanets acquire their mass?  There are many, often competing processes that regulate the accretion rate of protoplanets. To assess their effects we present a new, publicly-available toy model.  The rationale behind the toy model is that it encompasses as many physically-relevant processes as possible, but at the same time does not compromise its simplicity, speed, and physical insight.  The toy model follows a modular structure, where key features -- \eg\ planetesimal fragmentation, radial orbital decay, nebula turbulence -- can be \textit{switched} on or off.  Our model assumes three discrete components (fragments, planetesimals, and embryos) and is zero dimensional in space.  We have tested the outcomes of the toy model against literature results and generally find satisfactory agreement.  We include, for the first time, model features that capture the three-way interactions among small particles, gas, and protoplanets.  Collisions among planetesimals will result in fragmentation, transferring a substantial amount of the solid mass to small particles, which couple strongly to the gas.  Our results indicate that the efficiency of the accretion process then becomes very sensitive to the gas properties -- especially to the turbulent state and the magnitude of the disk headwind (the decrease of the orbital velocity of the gas with respect to Keplerian) -- as well as to the characteristic fragment size.

\end{abstract}


\section{Introduction}
\labelH{sec:Intro}
Giant, gas-rich planets are not formed in one day, but their formation should be  complete within the $\sim$10$^6$ yr over which the gas-rich phase of protoplanetary disks last \citep{FedeleEtal2010}.  As there are about 100 $e$-foldings required to grow a sufficiently massive core ($\sim$$10\ M_\oplus$; \citealt{PollackEtal1996}) out of ISM dust grains ($m\approx10^{-17}$ g), this nevertheless represents a fast process. Not only is the process fast: the planet formation mechanism also seems to be very efficient.  Over the years surveys like \textsc{Harps} and \textsc{Kepler} have discovered a wide number and variety of exoplanets \citep{MayorEtal2010,HowardEtal2011}.  Planet formation may be ubiquitous.

The view that the initial stage of giant planet formation proceeds similarly to that of terrestrial planet formation is commonly held.  This is the core accretion paradigm \citep{MizunoEtal1978,PollackEtal1996}: a rocky core is formed first, whereafter it binds the nebular gas, when it becomes sufficiently massive.  Another paradigm for the formation of massive planets is the disk-instability model, where the gaseous disk is sufficiently massive to become gravitationally unstable \citep[\eg][]{MayerEtal2007,Boley2009,Boss2011}.  Here, we will consider the core accretion paradigm and focus exclusively on the problem of forming a $\sim$$10\ M_\oplus$ core quickly enough.

An important intermediate step is the formation of planetesimals -- gravitationally interacting bodies usually thought to be of km-size or larger \citep{Safronov1969}.  These are the building blocks for planets.  Indeed, a typical assumption in protoplanet growth studies is that the majority of solids (\ie\ the rock and ice) of the protoplanetary disk resides in planetesimals.  But how planetesimals form in the first place remains ill-understood. Models favoring direct sticking by dust particles \citep{Weidenschilling1997}, have recently been confronted with a variety of obstacles: the fragmentation barrier at m-sizes \citep[\eg][]{BirnstielEtal2010i}; the bouncing barrier at mm-sizes \citep{GuettlerEtal2010,ZsomEtal2010,ZsomEtal2011}; or a charging barrier for even smaller (and fluffier) particles \citep{Okuzumi2009}.  Perhaps these results indicate that planetesimal formation requires special conditions: \eg\ sticky material properties (like ice; \citealt{WadaEtal2009}); particle pileups in drift-free regions \citep{YoudinChiang2004,KretkeLin2007,BrauerEtal2008i}; concentration of chondrule-size particles and their subsequent gravitational collapse \citep{CuzziEtal2010,Chambers2010}; or the shearing instability that operates on a dense layer of dm-m size boulders \citep{YoudinGoodman2005,JohansenEtal2007,JohansenEtal2009}.

The transition to planetesimal size bodies, instigates the runaway and oligarchic growth phases \citep{IdaMakino1993,KokuboIda1996,KokuboIda1998,OrmelEtal2010}.  Runaway growth is fast and a protoplanet seed forms \citep{WetherillStewart1989}.  However, at some point the runaway body faces a backlash from its own viscous stirring, thereby allowing embryos\footnote{We will employ the phrasing `oligarchs', `core', `embryo', and `protoplanet' mostly as synonyms.  More correctly, an embryo or protoplanet consist of a rocky (or icy) core with a gaseous atmosphere.} in neighboring zones to catch up in terms of mass.  What follows is a two component system of embryos and planetesimals, with the latter containing most of the solid mass.  The core formation challenge then amounts to transferring the mass reservoir of planetesimals onto these embryos.

There have been recent reviews of the runaway/oligarchic growth stage \citep[\eg][]{GoldreichEtal2004,LevisonEtal2010,OrmelEtal2010i} and we will not repeat these here, but the underlying problem is that during oligarchic growth embryos dynamically stir the planetesimals faster than they can accrete them.  For this reason, the next $e$-folding in embryo mass always takes longer than the previous. Allowing for embryo atmospheres \citep{InabaIkoma2003}, which enhances the radius at which planetesimals will be captured, alleviates the problem somewhat, but generally modelers require massive disks -- more massive than the minimum mass solar nebula (MMSN; \citealt{Weidenschilling1977i,HayashiEtal1985}) benchmark -- to form massive protoplanets within the time that the gas disk disperses.

\begin{figure*}[t]
  \plotone{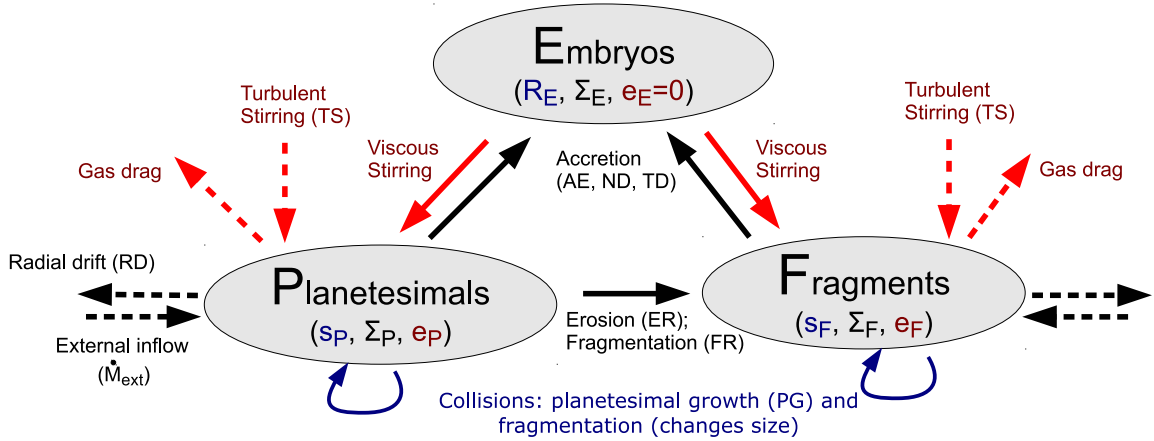}
  \caption{Sketch of processes affecting the dynamical state (\textit{red} arrows) and the surface density (\textit{black} arrows) of the components. Each of the three components is characterized by a surface density $\Sigma_k$, size, $s_k$ (or $R_E$ for the embryo's), and excitation state $e_k$ (the embryos' eccentricity is by definition 0).  \textit{Dashed} arrows denote external processes like gas drag and drift motions.  The random motion (eccentricity) of the planetesimal and fragment population is increased by viscous stirring and turbulent stirring, and damped by gas drag.  The surface density in embryos increases due to accretion.   The surface density in fragments and planetesimals changes to accretion, mutual collisions (erosion and fragmentation), radial drift, and inflow of solids from external regions.  Collisions within the planetesimal and fragment population changes the characteristic size ($s_P$ and $s_F$) of these bodies (\textit{blue} arrows).}
  \labelH{fig:sketch}
\end{figure*}
However, planetesimal-planetesimal collisions cause a copious amount of debris, especially for massive disks.  The question is whether these fragments -- or small particles in general -- are more conducive to growth than their km-size progenitors.  This is not an easy question to answer since smaller particles interact more strongly, and in more diverse ways, with the gas.  On the positive side, gas damping is much stronger, reducing their eccentricities.  Similarly, the dissipative nature of gas drag enhances capture rates.  However, orbital decay timescales can be extremely short \citep{Weidenschilling1977} and diffusion by turbulence could severely dilute their midplane densities.  One of the foci of this work is to capture these key physical processes into a common framework.

The two component approximation with which oligarchic growth is characterized renders it ideal to be studied by simple numerical or even analytical models \citep{GoldreichEtal2004,Chambers2006,Chambers2008,BruniniBenvenuto2008,GuileraEtal2011}. With a `toy model' (TM) we mean anything where the mass distribution is approximated by two or three (in case fragments are included) components.  This in contrast to models that include the full mass spectrum \citep[\eg][]{Weidenschilling1997,KobayashiEtal2010,BromleyKenyon2011i}, which may be referred to as `sophisticated'.  The goal of a TM may be to provide physical insight or to allow for a quick (but still sufficiently accurate) computation.  Population synthesis models, for example, \citet{IdaLin2004}, \citet{MordasiniEtal2009}, and their sequels employ a TM for the embryo growth stage, if not for all stages.  But do TMs capture all relevant physics?  There is often a tendency for the (above) TM to include more sophisticated features, \eg\ a radial dimension to follow the migration of protoplanets or multiple mass bins to capture the collisional cascade.

In this work we will attempt to adhere more strictly to what we consider characterizes a TM. Specifically this demands the model to be:
\begin{enumerate}
  \item Quick.  CPU speeds should be seconds, rather than hours, let alone days.
  \item General. That is, it should include all relevant physical process that could potentially affect protoplanet growth;
  \item Transparent. It must be clear which effect causes what.
\end{enumerate}
Combining the second and third properties is challenging, since a more complete model, characterized by more features, is often also a more complex one.  For that reason, we have opted for a modular approach, where features can be \textit{switched} on or off.  As a result of the first and third property, our TM contains many simplification, perhaps \textit{over}simplifications.  However, in our view, a TM is complementary, not in competition, to more sophisticated models of planet formation.  That is, the TM provides a first glance of what result may be expected.  It can explore the parameter space most efficiently and assesses whether or not novel features do matter.  It may not get the details correct; its results may even be off by several factors of unities.  But here it is where it can work in conjunction with sophisticated models.

In this paper, the emphasis lies on the presentation and validation of the toy model to follow protoplanet growth.  In \se{toy-model} we present the toy model.  In \se{tests} we validate it against several literature studies. In \se{applic} the novel features of the toy model are explored.  We summarize and conclude in \se{summ}.

\section{The toy model}
\labelH{sec:toy-model}
This section presents the toy model in detail.\footnote{The Python source code of the toy model is publicly available at \texttt{http://astro.berkeley.edu/$\sim$ormel/software.html}.}  In \se{overview} we give an overview of the features. In \se{accretion} we describe how the accretion rate, or , equivalently, the embryo growth timescale, is computed.  \Ses{dyn-state}{eh-sol} address how the dynamical state (eccentricities of planetesimal and fragments) is obtained.  \Se{fragmt} discusses collisional fragmentation, \ie\ the way collisions affect the surface density in planetesimal and fragments and their characteristic sizes.  \Se{radial} discusses how drift motions affect the surface density of the components.  We summarize the algorithm in \se{implement}.

\subsection{Overview, model switches, and caveats}
\labelH{sec:overview}
\begin{deluxetable*}{llp{10cm}l}
  \tablecaption{\labelH{tab:switches}The list of binary switches (in alphabetic order) that can be turned on or off.}
  \tablehead{
    Switch   & Abbr                     & Accounts for the \dots & Discussed in
    }
  \startdata
    Atmosphere enhancement              & \texttt{AE} &  Increase in the capture radius due to trapping of particles in protoplanets atmospheres. & \se{atmos-enh}\\
    Erosive collisions                  & \texttt{ER} &  (Fragmenting) collisions between fragments and planetesimals   & \se{eros}\\
    Fragmentation                       & \texttt{FR} &  Fragmentation of planetesimals by planetesimal-planetesimal collisions        & \se{fragmt}\\
    Oligarchy                           & \texttt{OL} &  Presence of neighboring protoplanets  & \ses{key-defs}{radial} \\
    Planetesimal growth                 & \texttt{PG} &  Growth of characteristic planetesimal size, due to planetesimal-planetesimal collisions  & \se{s-change}\\
    Nebular drag                        & \texttt{ND} &  The nebular flow in calculating the accretion rates of particles coupled to the gas. & \se{sd-reg}\\
    Radial drift                        & \texttt{RD} &  Depletion of the surface density due to orbital decay of planetesimals and fragments   & \se{radial}\\
    Turbulent diffusion                  & \texttt{TD} &  Vertical diffusion of particles due to turbulence, reducing their midplane density   & \se{sd-reg}\\
    Turbulent stirring                  & \texttt{TS} &  Additional excitation of planetesimals due to gravitational scattering by turbulence-induced density fluctuations & \se{turb-stir}
  \enddata
\end{deluxetable*}
A sketch of the toy model is given in \fg{sketch}. The mass distribution is approximated by three component: embryos (E), planetesimals (P) and fragments (F), each characterized by a size ($s_k$ or $R_E$ for the embryos), surface density $\Sigma_k$, and eccentricity $e_k$ where $k$ is one of E, P, or F.  These quantities are referred to as the state variables of the system.  In our TM we quantify how the components influence each other.  Embryos will dynamically excite the planetesimal and fragment population and their mass increases due to accretion.  Planetesimal fragmentation transfers mass from the planetesimal to the fragment population.  Collisions among these components can alter their characteristic size ($s_P$ or $s_F$).  Finally, the gas from the disk plays a crucial role: it damps the random motions (eccentricities) of planetesimals and fragments, but may also excite eccentricities due to turbulence-induced density inhomogeneities in the gas.  Furthermore, the gas depletes the surface density due to radial drift.  To complement the model we allow (optionally) for an external inflow of matter at a rate $\dot{M}_\mathrm{ext}$.

An overview of the binary switches the toy model contains is provided in \Tb{switches}.  These can be turned on or off dependent on the problem under consideration.  Due to this modular nature, they are referred to as \textit{switches}.  When a switch equals 0 it means the feature is turned off; when \texttt{switch=1}, it is turned on.  The atmosphere enhancement (\texttt{AE}) and nebula drag (\texttt{ND}) switches affect the efficiency of accretion.  Atmosphere enhancement results in a larger accretion cross section due to energy dissipation effects.  Likewise, the nebular gas causes the orbit to be modified from the usual (dissipationless) 2-body approximation, thereby also modifying the collision rate.   Fragmentation among planetesimals is included when \texttt{FR=1}, and planetesimal-fragment erosive collisions are additionally included when \texttt{ER=1}.  The oligarchic switch (\texttt{OL}) indicates the presence of neighboring protoplanets, which affects the accretion rate and the drift timescales of solids.  Growth of planetesimals is taken into account when \texttt{PG=1}; otherwise the planetesimals characteristic size $s_P$ does not evolve.  Radial drift (by fragments and planetesimals) is included when \texttt{RD=1}.  Stirring due to turbulence-induced density fluctuations is included when \texttt{TS=1}; otherwise, viscous stirring by embryos is the only mechanism that excites the eccentricities of the planetesimal bodies.  Finally, we consider the possibility of vertical diffusion of particles due to turbulent diffusion.  The \texttt{TD} switch is then turned on.

\begin{deluxetable*}{llp{10cm}l}
  \tablecaption{\labelH{tab:constants}List of parameters and frequently used symbols.}
  \tablehead{
  Symbol\tablenotemark{a} & Default value\tablenotemark{b} & Description  & Reference
  }
  \startdata
  $\Delta v_{kl}$         &                               & Relative velocity between particles of component $k$ and $l$                                    & \Se{dd} \\
$\Sigma_{[g,k]}$      &                                 & Gas surface density ($\Sigma_g$); solid surface density of component $k$ ($\Sigma_k$)      & \Eq{mmsn2} \\
$\Sigma_\mathrm{ini}$   & (mmsn)                        & Initial total surface density in solids                                                               & \Eq{mmsn2}\\
$\Omega$                & (mmsn)                        & Local orbital frequency corresponding to disk radius $a_0$                                      & \Se{key-defs} \\
$\alpha_E$              &                               & Embryo radius in terms of Hill radius                                                           & \Eq{alphaE} \\
$\alpha_\mathrm{ss}$    & $10^{-4}$                     & \citet{ShakuraSunyaev1973} turbulence viscosity parameter                                       & \Se{sd-reg} \\
$\gamma$                & 1.4                           & Adiabatic constant of the gas                                                                   & \Se{atmos-enh} \\
$\xi$                   & 0.1                           & Control parameter for timestep increment                                                        & \Se{implement} \\
$\kappa$                &0.01 $\mathrm{cm^2\ g^{-1}}$   & Atmosphere opacity                                                                              & \Se{atmos-enh} \\
$\rho$                  &                               & Gas density within the embryo's atmosphere (Appendix)                                           & \app{atmos} \\
$\rho_{[c,s]}$          & 1 $\mathrm{g\ cm^{-3}}$       & Internal density of the embryo's core ($\rho_c$); and solids (planetesimals, fragments; $\rho_s$ ) \\
$\rho_g$                & (mmsn)                        & Gas density of the nebula midplane at a disk radius $a_0$                                       & \Se{tests} \\
$\sigma_{kl}$           &                               & Cross section for collisions between components $k$ and $l$ (main paper); dimensionless density (Appendix)                        & \se{dd}; \app{atmos} \\
$\tau_\mathrm{fr}$      & (mmsn)                        & Dimensionless friction time ($=T_\mathrm{fr}\Omega$)                                            & \Ses{sd-reg}{gasfric} \\
$C_\mathrm{acc}$        & 1 or 1.5                      & Accretion rate correction factor due to embryo-embryo merging                                   & \Se{key-defs} \\
$C_\mathrm{drift}$      & 0.5                           & Prefactor for determining the drift timescale, applicable when \texttt{OL=1}                    & \Se{radial}, \eq{vdr-ol}   \\
$C_\mathrm{col}$        & 36                            & Numerical constant that enters the stirring rate expression                                         & \Eq{Pcol} \\ 
$C_\mathrm{gg}$         & 9.0                           & Prefactor in strength expression for gravitationally-bound bodies                               & \Eq{Qd} \\
$H_g$                   &                               & Scaleheight gas disk ($=c_s/\Omega$)                                                          & \Se{tests} \\
$L_\star$               & $1\ L_\odot$                  & Stellar luminosity                                                                              & \Eq{mmsn3} \\
$M_\star$               & $1\ M_\odot$                  & Stellar mass \\
$M_{[c,E]}$             &                               & [core, Embryo] mass\tablenotemark{c}  \\
$M_\mathrm{ini,E}$      & $10^{-6}\ M_\oplus$           & Initial core mass                                                                               & \Se{tests} \\
$M_\mathrm{iso}$       &                                & Isolation mass                                                                               & \Eq{Miso} \\
$\dot{M}_\mathrm{ext}(t)$& 0                            & Mass accretion rate of solids from exterior regions drifting past the embryo                    & \Se{radial} \\
$P_\mathrm{col}$        &                               & Specific collision rate in terms of Hill units                                                  & \Se{key-defs} \\
$P_\mathrm{col,2d}$     &                               & Specific collision rate in Hill units in the 2-dimensional limit                                & \Se{sd-reg} \\
$P_\mathrm{vs}$         &                               & Viscous stirring rate in Hill units                                                             & \Eq{deh2dt} \\
$P_\mathrm{vs,sd}$      & 73                            & Max stirring rate, obtained at Hill eccentricities $e_h\ll1$                                       & \Se{sd-reg} \\
$R_{[a,b,c,h,E]}$       &                               & [Atmosphere-enhanced, Bondi, core, Hill, embryo] radius\tablenotemark{c}                        & \Ses{key-defs}{atmos-enh} \\
$T\sub{col}{kk}$        &                               & Collision timescale for planetesimal-planetesimal or fragment-fragment collisions               & \Eq{Tcol} \\
$T\sub{[\Sigma,dr,fr]}{k}$  &                           & [Depletion, drift, friction] timescale of planetesimals or fragments                            & \Eqss{TsigF}{tdrag}{Tdrift} \\
$T\sub{gr}{k}$          &                               & Growth timescale of embryos due to accretion of component $k$                                   & \Eq{Tgr} \\
$T_g$        & (mmsn)                        & Temperature nebula at $a=a_0$                                                                   & \Eq{mmsn3} \\
$T_\mathrm{stir}$       &                               & Net stirring timescale (maximum of $T_\mathrm{vs}$ and $T_\mathrm{ts}$)                         & \Eq{Tstir}\\
$T_\mathrm{[ts,vs]}$    &                               & Timescale for [turbulent,viscous] stirring                                                      & \Eqs{Tts}{Tvs} \\
$Q_\mathrm{0g}$         & $10^7\ \mathrm{cm^2\ s^{-2}}$ & Prefactor for the gravity regime in the $Q_d^\ast$ law                                          & \Eq{Qd} \\
$Q_\mathrm{0s}$         & 2.1 $\mathrm{cm^5\ s^{-2} g^{-1}}$ & Prefactor for the strength regime in the $Q_d^\ast$ law                                    & \Eq{Qd} \\
$Q_d^\ast$             &                               & Material strength                                                                               & \Eq{Qd} \\
$W_\mathrm{sim}$       &                               & Radial width of the simulated region                                                                  & \Se{radial} \\
$W_\mathrm{neb}$        &                               & Dimensionless parameter for the calculation of atmosphere structure                                 & \Eq{Wneb} \\
$a_0$                   & 5.0 AU                        & Disk radius (semi-major axis)                                                                   & \\
$\tilde{b}$             & 12.5                          & Separation of oligarchs in units of the Hill radius $R_h$                                       & \Se{vs} \\
$b_{[g,s]} $            & [1.19, $-0.45$]               & Exponent of [gravity, strength] term in $Q_d^\ast$ law                                          & \Eq{Qd} \\
$c_s$                   & (mmsn)                        & Isothermal sound speed                                                                          & \Se{tests}\\
$[e_k, e\sub{h}{k}]$    &                               & [Orbital, Hill eccentricity] of component $k$                                                   & \Se{key-defs} \\
$e_{h1}$                & 2.34                          & Transition Hill eccentricity for $P_\mathrm{vs}$ fit                                            & \Se{vs}, \Fg{Pvs} \\
$h_k$                   &                               & Scaleheight particle layer accounting for Keplerian inclination and turbulent diffusion & \Eq{hp} \\ 
$i_k$                   &                               & Orbital inclination                                                                             & \Se{dd} \\
$q_k$                   &                               & Ratio collision energy over collision strength                                                  & \Eq{qi} \\
$s_k$                   &                               & Radius planetesimals or fragments \\
$s_\mathrm{F,ini}$               &                               & Initial radius fragments                                                                        & \Fg{Qdstar} \\
$s_\mathrm{P,ini}$               & 10 km                         & Initial radius planetesimals  \\
$v_K$                   &                               & Keplerian (orbital) velocity corresponding to $a_0$                                             & \\
$v_\mathrm{esc}$        &                               & Escape velocity of the embryo                                                                   & \Se{dd} \\
$v_h$                   &                               & Hill velocity                                                                                   & \Se{key-defs} \\
$v_\mathrm{hw}$         & 54 m\ s                       & Headwind velocity experienced a by heavy body due to sub-Keplerian moving gas                     & \Se{gasfric}, \Eq{eta}\\
$v\sub{rd}{k}$          &                               & Radial drift velocity for particles of component $k$                                            & \Eqs{vr}{vr2} \\
  \enddata
  \tablenotetext{a}{The dummy index $k$ (one of E, P, or F) indicates one of the three components of the toy model (embryos, planetesimals, or fragments).  Occasionally, similar symbols are combined in one line using square brackets. }
  \tablenotetext{b}{A number indicates the default value of the parameter; `(mmsn)' implies that (by default) we follow the minimum-mass solar nebula model (\se{tests}).}
  \tablenotetext{c}{The atmosphere is assumed not to contribute significantly to the embryo's mass, \ie\ $M_E = M_c$.  The embryo radius $R_E$, as it appears throughout \se{accretion} is either the atmosphere capture radius $R_a$ (when \texttt{AE=1}) or the core radius $R_c$ (when \texttt{AE=0}). }
\end{deluxetable*}

Other (continuous) parameters of the toy model are listed in \Tb{constants}.  The toy model is further characterized by the following properties and assumptions:
\begin{enumerate}
  \item We prefer physically-motivated order of magnitude expressions rather than detailed black-box formulas.  This implies that we favor physical insights at the expense of precision.
  \item There is no explicit radial dimension. Type-I migration is neglected.  Removal of (small) particles by aerodynamic drift is included (if \texttt{RD=1}), and an (ad-hoc) injection rate of mass may be provided ($M_\mathrm{ext}\neq0$).  A global interpretation (see \se{radial}) can be made when oligarchy is assumed (\texttt{OL=1}), but fundamentally the model is strictly local.
  \item The model assumes a bimodal (or trimodal) distribution for the mass.  This implies an instantaneous jump from the planetesimal size $s_P$ to the fragment size $s_F$. Furthermore, the three-component assumption is not applicable to follow the runaway growth phase (the phase preceding oligarchic growth).
  \item Regarding the dynamical state (computation of eccentricities), the model assumes a balance between the (viscous) stirring timescale and a limiting timescale (\se{eh-sol}).  We do not integrate the stirring rates ($de^2_k/dt$ or $di^2_k/dt$).  This setup implies memoryless behavior; \eg\ the eccentricities follow from the current values of $\Sigma_k$ and $M_E$.
  \item We implicitly assume that $\Sigma_E \ll \Sigma_F+\Sigma_P$, such that the random motions of protoplanets are negligible due to dynamical friction ($e_E=0$).  The toy model is therefore only valid as long as oligarchy pertains.  
\end{enumerate}

\subsection{The accretion rate}
\labelH{sec:accretion}
\subsubsection{Hill units and key definitions}
\labelH{sec:key-defs}
We define the total accretion rate of the embryo as
\begin{equation}
  \frac{dM_E}{dt} = \sum_{k=\mathrm{P,F}} C_\mathrm{acc} P\sub{col}{k} \Sigma_k R_h v_h,
  \labelH{eq:dMdt}
\end{equation}
where $M_E$ is the embryo mass, $C_\mathrm{acc}$ a correction factor for oligarchic growth (discussed below), $\Sigma_k$ the surface density and the sum is over the planetesimals ($k=P$) and fragments ($k=F$) components.  The Hill radius $R_h$ is defined as
\begin{equation}
  R_h = a_0 \left( \frac{M_E}{3M_\star} \right)^{1/3},
  \labelH{eq:Rh}
\end{equation}
with $a_0$ the semi-major axis of the protoplanet, $M_\star$ the stellar mass, $v_h=R_h\Omega$ the Hill velocity, and $\Omega=\Omega(a_0)$ the orbital (Kepler) frequency at $a_0$.  Finally, $P_\mathrm{col}$ is the specific accretion rate, which depends on the eccentricities ($e$) of the particles, their scaleheight (or inclination, $i$), and the dimensionless radius of the embryo:
\begin{equation}
  \alpha_E 
  = \frac{R_E}{R_h}
  = \frac{1}{a_0} \left( \frac{9M_\star}{4\pi\rho_c} \right)^{1/3} 
  \approx 10^{-3} \left( \frac{5\ \mathrm{AU}}{a_0} \right) \left( \frac{\rho_c}{3\ \mathrm{g\ cm^{-3}}} \right)^{-1/3},
  \labelH{eq:alphaE}
\end{equation}
where $R_E$ is the radius of the embryo (the core radius) and $\rho_c$ its density.  Note that $\alpha_E$ is primarily a function of disk radius $a_0$.  We further introduce the Hill eccentricity $e_h = ev_K/v_h$.  The use of the Hill eccentricity $e_h$ is convenient since it distinguishes the dispersion-dominated (d.d.) regime ($e_h>1$), where approach velocities are determined by the planetesimal eccentricities, from the shear-dominated (s.d.) regime ($e_h<1$), where approach velocities are given by the shear of the Keplerian-rotating disk. 

In \eq{dMdt} $C_\mathrm{acc}$ represents a correction factor to $dM/dt$, relevant for oligarchic growth,  due to the self-accretion of embryos.  Its value can be obtained as follows.  During their growth embryos keep a distance of $\tilde{b}R_h$ with $\tilde{b}\approx12.5$ constant \citep{KokuboIda1998}.  Since $R_h\propto M^{1/3}$ the number of embryos has halved when their mass has increased by a factor 8.  Thus, the growth by a factor of 8 is split with embryo-embryo growth accounting for a factor of 2 and embryo-planetesimal growth accounting for a factor of 4. Therefore, $C_\mathrm{acc}=1.5$, when we consider oligarchic growth (\texttt{OL=1}).  Otherwise, if we consider a single protoplanet (\texttt{OL=0}) the correction does not apply and $C_\mathrm{acc}=1$.

From \eq{dMdt}, the accretion (or growth) timescale for each component is defined as
\begin{equation}
  T\sub{gr}{k} \equiv \frac{M_E}{dM_E/dt} = \frac{M_E}{C_\mathrm{acc}P\sub{col}{k}\Sigma_k R_h v_h}.
  \labelH{eq:Tgr}
\end{equation}
Next, we derive expressions for $P_\mathrm{col}$ for the d.d.-\ ($e_h>1$) and s.d.-regimes ($e_h<1$) .

\begin{figure*}[t]
  \plotone{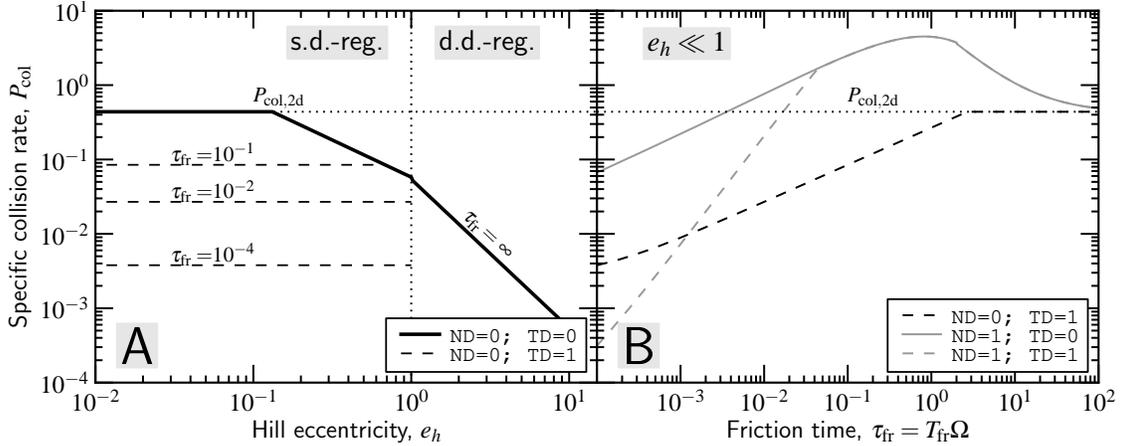}
  \caption{Specific collision rate $P_\mathrm{col}$ for a 0.1 $M_\oplus$ protoplanet situated at 5 AU. \textit{(left)} $P_\mathrm{col}$ as function of Hill eccentricity, $e_h$. For a gas-free system (\texttt{ND=TD=0}) $P_\mathrm{col}$ is a function of $e_h$ only (\textit{thick solid} line).  Adding turbulent diffusion (\texttt{TD=1} with $\alpha_\mathrm{ss}=10^{-4}$; \textit{dashed} lines) lowers the collision rate in the shear-dominated regime and renders it a function of the particle size or dimensionless stopping time $\tau_\mathrm{fr}=T_\mathrm{fr}\Omega$, rather than of $e_h$. \textit{(right)} $P_\mathrm{col}$ as function of $\tau_\mathrm{fr}$ in the s.d.-regime, valid for $e_h\ll1$ and a headwind of $v_\mathrm{hw}=54\ \mathrm{m\ s^{-1}}$.  The curves give $P_\mathrm{col}$ for the case including turbulent diffusion (\textit{dashed} curves) and for the modification of the collisional cross section by nebular drag (\textit{gray} curves; \texttt{ND=1}).\\}
  \labelH{fig:pcol}
\end{figure*}
\subsubsection{The dispersion-dominated regime: $e_h>1$}
\labelH{sec:dd}
In the dispersion-dominated (d.d.) regime the relative velocity between particle and protoplanet is given by the eccentricity of the former, $\Delta v_\mathrm{EP} = e_P v_K$ where $v_K = a_0 \Omega$ is the Kepler velocity.  The d.d.-regime is generally valid for big bodies, orbiting on a well defined Kepler orbit of inclination that is half of the eccentricity, $i=e/2$ -- the so called equilibrium value.  For $P_\mathrm{col}$ we adopt
\begin{equation}
  P_\mathrm{col} = C_\mathrm{col} \frac{\alpha_E}{e_h^2}
  \labelH{eq:Pcol}
\end{equation}
with $C_\mathrm{col}=36$ \citep{GreenzweigLissauer1990,GreenzweigLissauer1992}.

As an illustration, let us verify the validity of \eq{Pcol} by an order-of-magnitude argument.  The rate at which a protoplanet sweeps up particles can be written as
\begin{equation}
  \frac{dM_E}{dt} \approx n_P \sigma_\mathrm{EP} (\Delta v_\mathrm{EP}) m_P,
  \labelH{eq:dMdt-1}
\end{equation}
where $(n_P \sigma_\mathrm{EP} \Delta v_\mathrm{EP})^{-1}$ is the collision timescale between planetesimals and embryos, $m_P$ the mass of the planetesimal, $n_P$ the number density of planetesimals, $\sigma_\mathrm{EP}$ the gravitationally-enlarged collision cross section.  Gravitational focusing enlarges the collisional cross section to $\sigma_\mathrm{EP} = \pi R_E^2 (v_\mathrm{esc}/\Delta v_\mathrm{EP})^2$ where $v_\mathrm{esc}=\sqrt{2GM_E/R_E}$ is the escape velocity of the embryo with $G$ Newton's gravitational constant.\footnote{More correctly, the focusing factor is $1 + v_\mathrm{esc}^2/(\Delta v_\mathrm{EP})^2$ but in our case we ignore the unity term since gravitational focusing factors for oligarchic growth are always $\gg$1.}  Furthermore, we can write $n_P = \Sigma_P/2h_P m_P$, with $h_P\approx ia_0$ the scaleheight of the planetesimal layer and $i=e/2$ the inclination.  Then, \eq{dMdt-1} becomes
\begin{equation}
  \frac{dM_E}{dt} \approx  \pi R_E^2 \Sigma_k \Omega \left( \frac{v_\mathrm{esc}}{ev_K} \right)^2.
  \labelH{eq:dMdt-2}
\end{equation}
It can be shown that $R_Ev_\mathrm{esc}^2 = 6R_h v_h^2$; and since $e_h v_h = ev_K$ the above equation, up to a constant, equals \eq{dMdt} with \eq{Pcol} for $P_\mathrm{col}$.  The larger value of $C_\mathrm{col}$ above is due to the averaging over a distribution of eccentricities.

\subsubsection{The shear-dominated regime: $e_h<1$}
\labelH{sec:sd-reg}
In the shear-dominated (s.d.)\ regime -- valid mostly for smaller particles -- the expression for $P_\mathrm{col}$ is not so straightforward as \eq{Pcol}.  First, inclinations start to decouple from the eccentricities \citep{IdaEtal1993}; the ratio $i/e=1/2$ is no longer maintained.  Nevertheless, we will assume this relation to hold as we do not separately compute the inclinations.  More important is the way the gas interacts with the (small) fragments -- a subtlety that is sometimes neglected in studies of oligarchic growth.  We consider two effects:
\begin{enumerate}
  \item When disks are turbulent, diffusion of particles causes them to spread over a vertical height $h_k$ that may be much larger than the corresponding value due to their inclination ($ia_0$).  Here, we take the ensuing scaleheight for the dust as
\begin{equation}
  \labelH{eq:Hdust}
  H_g\sqrt{\frac{\alpha_\mathrm{ss}}{\alpha_\mathrm{ss} + \tau_\mathrm{fr}}}
\end{equation}
\citep{YoudinLithwick2007} where $\tau_\mathrm{fr} \equiv T_\mathrm{fr}\Omega$ is the dimensionless stopping time (see definition in \se{gasfric}), $\alpha_\mathrm{ss}$ the familiar turbulent diffusion parameter first introduced by \citet{ShakuraSunyaev1973}, and $H_g$ the scaleheight of the gas. The net scaleheight of the particle layer -- the quantity used in $P_\mathrm{col}$ (see below) -- is approximated as:
\begin{equation}
  h_P = \max \left( ia_0, H_g\sqrt{\frac{\alpha_\mathrm{ss}}{\alpha_\mathrm{ss} + \tau_\mathrm{fr}}} \right). \qquad (\texttt{TD=1})
  \labelH{eq:hp}
\end{equation}
Thus, when $\tau_\mathrm{fr}\ll \alpha_\mathrm{ss}$ the particles' scaleheight becomes equal to that of the gas.  But even when $\tau_\mathrm{fr}>\alpha_\mathrm{ss}$ turbulent diffusion can result in a scaleheight much larger than that from inclination alone ($h_P=ia_0$ when $\alpha_\mathrm{ss}=0$ or \texttt{TD=0}). When gas-drag effects do not affect $P_\mathrm{col}$ (\ie\ \texttt{ND=0}), the maximum collision rate is attained when the particles reside in a thin layer ($h_P\rightarrow 0$). For this gas free, 2D configuration $P_\mathrm{col} = P_\mathrm{col,2d} \approx 11\alpha_E^{1/2}$ \citep[\eg][]{IdaNakazawa1989}.  This 2D limit is valid when the particle scaleheight is less than $b_\mathrm{col}$, the impact radius for accretion.  Otherwise, the local density of particles is diluted by a factor $b_\mathrm{col}/h_P$ and the collision rate reduces accordingly:
\begin{equation}
  P_\mathrm{col} = P_\mathrm{col,2d} \min \left( 1, \frac{b_\mathrm{col}}{h_P} \right),
  \labelH{eq:Pcol-gf}
\end{equation}
where in the s.d.-regime, $b_\mathrm{col} = 1.7\alpha_E^{1/2} R_h$ (see \eg\ \citealt{InabaEtal2001,OrmelKlahr2010}).

\item Gas drag forces cause an encounter to deviate from its 2-body, energy-conserving, trajectory.  Considering this effect (\texttt{ND=1}) gives rise to a $P_\mathrm{col}$ and $b_\mathrm{col}$ that are different from the gas-free case, discussed above.  \citet{OrmelKlahr2010} (see also \citealt{PeretsMurray-Clay2011}) have accounted for the behavior of the gas drag \textit{during} the encounter and have derived analytic expressions for $b_\mathrm{col}$ and $P_\mathrm{col}$ in the presence of gas drag.  Therefore, when the nebular drag switch is turned on (\texttt{ND=1}), $b_\mathrm{col}$ and $P_\mathrm{col,2d}$ follow from the \citet{OrmelKlahr2010} study.  These expressions are summarized in \app{OK10}.  If, in addition, vertical diffusion is considered (\texttt{TD=1}), the correction factor to $P_\mathrm{col,2d}$ (\eqp{Pcol-gf}) is also applied.
\end{enumerate}

In \fg{pcol} the effects of these features on $P_\mathrm{col}$ is illustrated for a $0.1\ M_\oplus$ protoplanet at $5$ AU of internal density $\rho_c=3\ \mathrm{g\ cm^{-3}}$. The thick solid line in \fg{pcol}a shows $P_\mathrm{col}$ for the gas-free case, which is equivalent to setting the switches \texttt{ND} and \texttt{TD} to zero.  $P_\mathrm{col}$ increases with decreasing eccentricity.  In the d.d.-regime ($e_h>1$) it is given by \eq{Pcol}.  The s.d.-regime splits into two regimes: \sumi\ the 2D limit corresponding to a thin particle layer ($e_h\ll1$; dotted horizontal line); and a transition regime, where the thickness of the particle layer is larger than the accretion radius, $ia_0>b_\mathrm{col}$. (Note that we assume $i=e/2$ throughout). Although our formulation is slightly different, the $P_\mathrm{col}$ we obtain for the gas-free limit is consistent with the expressions given by \citet{InabaEtal2001}.

Allowing (small) particles to diffuse by turbulence (\texttt{TD=1}) decreases their density at the midplane and decreases $P_\mathrm{col}$. The dashed lines in \fg{pcol}a show that the corresponding reduction, valid for $\alpha_\mathrm{ss}=10^{-4}$, can become quite significant for small particles.  The correction factor is only applied in the s.d.-regime, i.e., for $e_h<1$, since it only matters for small particles that anyway satisfy $e_h\ll1$.  For this reason, it is more useful to plot $P_\mathrm{col}$ as function of size, or, rather, dimensionless friction time, see \fg{pcol}b, black dashed line. With increasing $\tau_\mathrm{fr}$, $P_\mathrm{col}$ increases until the 2D-limit is reached (dotted horizontal line). 

Accounting for the nebular flow around the protoplanet (\texttt{ND=1}) results in accretion rates that are quite different from $P_\mathrm{col}$ for the gas free case. As outlined by \citet{OrmelKlahr2010} an important parameter in determining the accretion rates is the magnitude of the disk headwind $v_\mathrm{hw}$. The gas in the protoplanetary disk is partly pressure supported, resulting in a angular velocity slower than the gas by a quantity $v_\mathrm{hw} = \eta v_K$, where $\eta$ is defined as \citep{AdachiEtal1976}:
\begin{equation}
  \labelH{eq:eta}
  \eta(a_0)  = \frac{1}{2\rho_g a_0 \Omega^2} \frac{dP_g}{da} = -\frac{1}{2} \left( \frac{c_s}{v_K^2} \right)^2 \frac{\partial \ln P_g}{\partial \ln a},
\end{equation}
where $\rho_g$ is the gas density of the nebula at disk radius $a_0$, $P_g=\rho_gc_s^2$ the gas pressure, $c_s$ the isothermal sound speed, and the gradient is evaluated at $a_0$. A heavy body (planet or planetesimal) is not influenced by the pressure gradient and moves at the Keplerian velocity; it therefore experiences a headwind of magnitude $v_\mathrm{hw}=\eta v_K$.  The solid gray line in \fg{pcol}b gives $P_\mathrm{col}$ for a headwind of $v_\mathrm{hw}=54\ \mathrm{m\ s^{-1}}$ (see \app{OK10} for a summary how $P_\mathrm{col}$ is derived when \texttt{ND=1}).  For $\tau_\mathrm{fr}\gg1$ one recovers the 2D, gas-free limit (large bodies are not influenced by gas drag). Somewhat smaller particles ($\tau_\mathrm{fr}\gtrsim1$), however, have a finite probability to be captured during their passage through the Hill sphere.  This capture mechanism is independent of the physical size of the protoplanet.  If the protoplanet is sufficiently massive, it also operates for particles $\tau_\mathrm{fr}\lesssim1$.  However, for very small particles ($\tau_\mathrm{fr}\ll1$) the gas drag force is so strong that accretion is suppressed, because particles are virtually glued to the gas.  In that case, the accretion rate falls below that of the gas free-limit.  Including turbulent diffusion (dashed gray line) further decreases $P_\mathrm{col}$.

\subsubsection{Atmosphere enhancement (\texttt{AE}) of the collision radii}
\labelH{sec:atmos-enh}
The collision radii further increase when protoplanets acquire an atmosphere.  Since this is generally true, \ie\ for fragments as well as planetesimals, we discuss it separately.  However, its effects are most pronounced for small particles.

A large protoplanet will acquire a dense atmosphere, which enhances the effective capture radius for collisions.  \citet{InabaIkoma2003} and \citet{TanigawaOhtsuki2010} have studied this effect by calculating the energy loss of the planetesimals near the point of their closest approach to the planet (where the atmosphere is densest).  In this way, they derive an enhanced radius for accretion,  $R_a$.  To obtain the correct accretion rate, this `atmosphere radius' must be used instead of the core radius ($R_c$ or the dimensionless $\alpha_c$) in all of the above expressions.  This is the procedure we adopt here.  In any case, we cap $R_a$ when it reaches either the Hill radius or the Bondi radius, $R_b=GM/\gamma c_s^2$ with $\gamma=1.4$.

\citet{InabaIkoma2003} provide a solution for the density structure of the atmosphere.  Their solution is quite general; it allows for multiple shells, where the conditions regarding, \eg\ the opacity can be different.  Here, we will make the additional simplification that the opacity is low and constant, $\kappa=0.01\ \mathrm{cm^2\ g^{-1}}$, reflecting essentially a grain-free atmosphere.  A low opacity may find some justification from the fact that \sumi\ most of the mass is in macroscopic bodies (planetesimals); and \sumii\ grains may otherwise quickly coagulate \citep{MovshovitzEtal2010,HelledBodenheimer2011}.  Although, a constant opacity throughout the atmosphere is arguably a somewhat crude assumption, our simplified approach manages to obtain good correspondence with the full solutions of \citet{InabaIkoma2003}, see \app{atmos}.

By virtue of the low opacity, we assume that the atmosphere is always radiative.  In \app{atmos} we approximate the equations of stellar structure and obtain an explicit, closed-form solution for the density as function of depth, $\rho=\rho(r)$.  Using this solution we straightforwardly solve for the enhanced radius, $R_a$:  
\begin{equation}
  R_a \approx R_b \left\{ \begin{array}{ll}
    \displaystyle
    \left[ 1 +\frac{2W_\mathrm{neb} (\sigma_a-1) +\log \sigma_a}{\gamma} \right]^{-1}; & (1\le \sigma_a \le \sigma_1) \\
    \displaystyle
    \left[ \frac{1}{x_1} + \frac{4(4W_\mathrm{neb})^{1/3}}{\gamma}\left(\sigma_a^{1/3}-\sigma_1^{1/3}\right) \right]^{-1}; & (\sigma_a > \sigma_1) \\
  \end{array}
  \right.
  \labelH{eq:Ra-sol}
\end{equation}
where $\sigma_a = \rho_a/\rho_g$ is the ratio of the density required to capture the particle and the nebula gas density (see \app{atmos}), $\sigma_1 = 1/5W_\mathrm{neb}$ a dimensionless density threshold that specifies the transition from the nearly isothermal regime to the pressure supported regime, and $x_1=R_1/R_b= 1+2W_\mathrm{neb}(\sigma_1-1) + \log \sigma_1$ the dimensionless radius corresponding to $R_1$.  The parameter that sets the structure of the atmosphere is $W_\mathrm{neb}$:
\begin{eqnarray}
  \labelH{eq:Wneb}
  W_\mathrm{neb} &=&  \frac{3\kappa L_c}{64\pi\sigma_\mathrm{sb}} \frac{P_g}{GM_c T_g^4}
  =  \frac{k_B \kappa \rho_g \rho_c R_c^2}{16\sigma_\mathrm{sb} \mu T_g^3 T_\mathrm{gr}} \\
  &\approx& 10^{-5} \left( \frac{R_c}{10^3\ \mathrm{km}} \right)^2 \frac{\kappa}{10^{-2}\ \mathrm{cm^2\ g^{-1}}} \\
  && \times \frac{\rho_c}{\mathrm{g\ cm^{-3}}} \frac{\rho_g}{10^{-10}\ \mathrm{g\ cm^{-3}}} \left( \frac{T_g}{\mathrm{100\ K}} \right)^{-3} \left( \frac{T_\mathrm{gr}}{\mathrm{Myr}} \right)^{-1},
  \nonumber
\end{eqnarray}
where $k_B$ is Boltzmann's constant, $\mu=2.34$ the mean molecular mass, $\sigma_\mathrm{sb}$ Stefan-Boltzmann constant, $P_g=kT_g\rho/\mu$ the pressure of the nebula, and $T_g$ the nebular temperature.  For the luminosity due to planetesimal accretion $L_c$ we have substituted $L_c = (GM_c/R_c) dM_c/dt = GM_c^2/R_c T_\mathrm{gr}$,  with $T_\mathrm{gr}$ the growth timescale and $R_c$ the core radius.  Our solution for $R_a$ (\eqp{Ra-sol}) requires that $W_\mathrm{neb}\ll1$.

Admittedly, a certain level of ambiguity is present between the atmosphere-enhanced radius $R_a$ and the enhanced collision radius, $b_\mathrm{col}$, as obtained from \citet{OrmelKlahr2010}, due to nebular gas-drag effects.  Both studies account for gas drag;  but where \citet{OrmelKlahr2010} consider the change (enhancement \textit{or} decrease) of $P_\mathrm{col}$ to originate from the interaction with the \textit{nebular gas flow} (assumed to have a constant density), \citet{InabaIkoma2003} calculate the enhancement of $P_\mathrm{col}$ due to the \textit{increase in density} in the direct vicinity of the planet, \ie\ its atmosphere, while ignoring the flow pattern.  Indeed, particles strongly coupled to the gas flow may be inhibited to accrete \citep{SekiyaTakeda2003,Paardekooper2007,OrmelKlahr2010}. Further work must account simultaneously for both the gas flow and the density structure of the planet -- like in \citet{LissauerEtal2009}, but for a wide parameter range.

In this study, we will merge the \citet{InabaIkoma2003} and \citet{OrmelKlahr2010} approaches.  Specifically,  we first calculate the atmosphere radius $R_a$ (\eqp{Ra-sol}), which then replaces the embryo radius $R_E$ in all the above expressions involving $P_\mathrm{col}$.  That means that \eq{alphaE} is superseded by $\alpha_E \equiv R_a/R_h$. For the protoplanetary growth stage, the atmosphere does not contribute much to the total mass; \ie\ the embryo mass $M_E$ remains dominated by the mass of the solid core $M_c$. 

\subsection{Stirring, friction, and drift timescales}
\labelH{sec:dyn-state}
\subsubsection{Viscous stirring}
\labelH{sec:vs}
Similar to the accretion rate, \eq{dMdt}, we define the stirring rate for (planetesimal) bodies also in terms of Hill units:
\begin{equation}
  \frac{de_h^2}{dt} \equiv P_\mathrm{vs} N_E R_h v_h
  \labelH{eq:deh2dt}
\end{equation}
where $N_E$ is the column density of embryos and $P_\mathrm{vs}$ the specific viscous stirring rate, which is function of eccentricity ($e_h$).  
\begin{figure}[t]
  \plotone{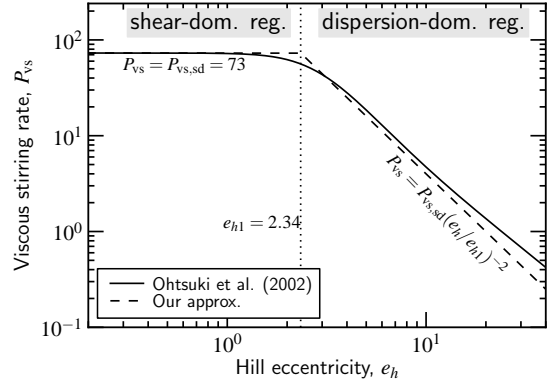}
  \caption{Comparison between the adopted stirring curve of our toy model (\textit{dashed} line) to the fit of \citet{OhtsukiEtal2002} (\textit{solid} curve). A 1:2 ratio  between inclinations and eccentricities is assumed. Our two-segment approximation breaks at an eccentricity $e_h=e_{h1}=2.34$.}
  \labelH{fig:Pvs}
\end{figure}
\Fg{Pvs} shows the dimensionless viscous stirring rate $P_\mathrm{vs}$ \citep{OhtsukiEtal2002} for the case that inclinations and eccentricities have reached a so called equilibrium ($i/e\approx0.5$; \citealt{IdaEtal1993}).  The basic features -- a plateau for $e_h\ll 1$ and a $e_h^{-2}$ decline for $e_h\gg1$ -- can be understood from geometrical arguments \citep{Ida1990,GoldreichEtal2004,OrmelEtal2010i}, but the detailed functional form of $P_\mathrm{vs}(e_h)$ follows from calibration against $N$-body experiments \citep{StewartIda2000,OhtsukiEtal2002}.  Here, we simplify matters by adopting a simple 2-segment fit-by-eye to the \citet{OhtsukiEtal2002} curve (which is itself a fit), accurate enough for our level of precision.  See the dashed line in \fg{Pvs}.

Using \eq{deh2dt}, the viscous stirring timescale is
\begin{equation}
  T_\mathrm{vs} 
  \equiv \frac{e_h}{de_h/dt} 
  = \frac{2e_h^2}{de_h^2/dt} 
  = \frac{4\pi \tilde{b} e_h^2 a_0}{P_\mathrm{vs} R_h} \Omega^{-1},
  \labelH{eq:Tvs}
\end{equation}
where we have inserted $N_E = 1/(2\pi a_0 \tilde{b} R_h)$.  That is, each protoplanet is assumed to stir an annulus of width $\tilde{b}$ times its Hill radius of the protoplanet, where $\tilde{b}$ is a parameter whose default value is $\tilde{b}=12.5$ \citep{KokuboIda1998}.  The stirring timescale is plotted by black lines in \fg{Tvs} for conditions typical of 5 AU. 

\subsubsection{Turbulent stirring}
\labelH{sec:turb-stir}
\begin{figure}[t]
  \centering
  \plotone{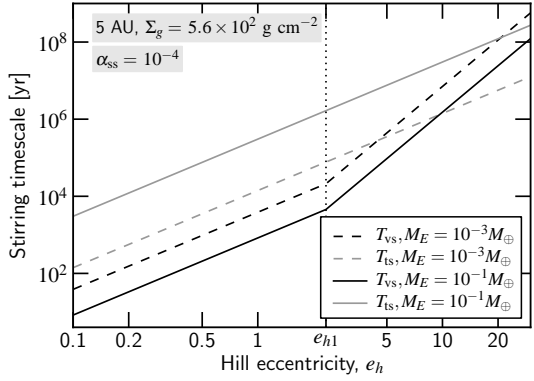}
  \caption{Viscous (\textit{black} lines) and turbulent (\textit{gray} lines) stirring timescales for protoplanets of 0.1 and $10^{-3}\ M_\oplus$ at a distance of $5 \mathrm{AU}$.}
  \labelH{fig:Tvs}
\end{figure}
Apart from stirring by protoplanets, we optionally include stirring of planetesimals by turbulent-induced density fluctuations in the gas.  This `turbulent stirring' (TS) has recently been explored by (among others) \citet{LaughlinEtal2004,IdaEtal2008,BaruteauLin2010,YangEtal2009,GresselEtal2011}.  We follow here a prescription by \citet{BaruteauLin2010}, which connects the intensity of the gravitational torques that the fluctuations induce to the turbulent $\alpha_\mathrm{ss}$ parameter.  This results in a turbulent stirring timescale of (see \app{densflucstir}):
\begin{equation}
  T_\mathrm{ts} = \frac{5\times10^{-3} e^2}{\alpha_\mathrm{ss}} \left( \frac{H_g a_0 \Sigma_g}{M_\star} \right)^{-2} \Omega^{-1}.
  \labelH{eq:Tts}
\end{equation}
Turbulent stirring is most important for small protoplanets and massive turbulent disks (see \fg{Tvs}).  Otherwise, viscous stirring dominates (see \app{densflucstir} and \fg{Tvs}).  

When turbulent stirring is implemented, we compute both $T_\mathrm{vs}$ and $T_\mathrm{ts}$. Otherwise, we only use viscous stirring, \ie
\begin{equation}
  \label{eq:Tstir}
  T_\mathrm{stir} =  \left\{ \begin{array}{ll}
    T_\mathrm{vs}                         & (\texttt{TS=0}) \\
    \min (T_\mathrm{vs}; T_\mathrm{ts})   & (\texttt{TS=1})
  \end{array} \right.
\end{equation}

\subsubsection{Gas friction and radial drift}
\labelH{sec:gasfric}
The stopping or friction time -- the time over which the particle velocity is changed by friction with the gas -- is defined as:
\begin{equation}
  \labelH{eq:tdrag}
  T\sub{fr}{k}(v_\mathrm{gas}) = \left\{ \begin{array}{lll}
    \displaystyle
    \frac{8s_k\rho_s}{3C_d \rho_g v\sub{gas}{k}}     &  (s_k>\frac{9}{4}l_\mathrm{mfp}); \\[5mm]
    \displaystyle
    \frac{s_k\rho_s}{v_\mathrm{th} \rho_g}            & (s_k<\frac{9}{4}l_\mathrm{mfp}); \\
  \end{array} \right.
\end{equation}
($k=$ P, F) where $s_k$ is the radius of the particle, $\rho_s$ its internal density, $l_\mathrm{mfp}$ the mean free path of the gas, $C_d$ the drag constant, which is a function of particle size, $v\sub{gas}{k}$ the gas-particle relative velocity, and $v_\mathrm{th}$ the thermal speed.  For the drag constant we follow \citet{Weidenschilling1977} and use
\begin{equation}
  \labelH{eq:Cd}
  C_\mathrm{ss} = \left\{ \begin{array}{lll}
    24 Re_p^{-1}    & \textrm{if} & Re_p<1 \\
    24 Re_p^{-0.6}\quad  & \textrm{if} & 1 \le Re_p \le 800 \\
    0.44            & \textrm{if} & Re_p > 800 \\
  \end{array} \right.
\end{equation}
with $Re_p = 2sv_\mathrm{gas}/\nu_\mathrm{mol}$ the particle Reynolds number with $\nu_\mathrm{mol}= l_\mathrm{mfp} v_\mathrm{th}/2$ the molecular viscosity of the gas.  

Due to the disk headwind $v_\mathrm{hw}$ (see \eqp{eta}), solids will lose angular momentum. This causes a particle to spiral in at a radial velocity \citep[\eg][]{Weidenschilling1977,NakagawaEtal1986,BrauerEtal2007}:
\begin{equation}
  v_r = - \frac{2\tau_\mathrm{fr}}{1+\tau_\mathrm{fr}^2} v_\mathrm{hw}.
  \labelH{eq:vr}
\end{equation}
The above expression holds for small particles: they will pile up in pressure maximums where $v_\mathrm{hw}=0$.  \Eq{vr} is zero if the headwind vanishes -- but this does not fully capture the behavior of heavy bodies on eccentric orbits. To first order in eccentricity (the epicyclic approximation), they feel a headwind 50\% of the time and a tailwind for the other 50\%, such that the net torque the gas drag exerts is zero.  However, to second order in $e$ there is a net effect: angular momentum is removed from the body, causing it to drift in.  This effect amounts to the planetesimal experiencing a net headwind of $\approx$$e^2 v_K$ and we write, more generally than \eq{vr} 
\begin{equation}
  v_\mathrm{rd} = - \frac{\tau_\mathrm{fr}}{1+\tau_\mathrm{fr}^2} (2v_\mathrm{hw} +e^2v_K).
  \labelH{eq:vr2}
\end{equation}
(Within order of unity, this simple formula captures the detailed orbit-averaging calculations performed by \citealt{AdachiEtal1976}.)

The particle-gas velocity $v_\mathrm{gas}$ is constructed in a similar vein. It consist of a part due to the inclinations and eccentricities of the bodies ($\approx$$ev_K$) and a contribution from the slower than Keplerian moving gas. The latter is due to the radial drift (\eqp{vr}) when $\tau_\mathrm{fr} \ll1$ and equals $v_\mathrm{hw}$ (the azimuthal headwind) when $\tau_\mathrm{fr}\gg1$.  Therefore we have approximately
\begin{equation}
  v_\mathrm{gas} \approx ev_K +v_\mathrm{hw} \frac{2\tau_\mathrm{fr}+\tau_\mathrm{fr}^2}{1+\tau_\mathrm{fr}^2}.
  \labelH{eq:vgas}
\end{equation}
The friction time can be found using an iterative procedure involving \eqs{tdrag}{vgas}.

\subsection{Solving for $e_h$}
\labelH{sec:eh-sol}
The adopted power-law approximations for the stirring timescale readily allow us to solve for the Hill eccentricity assuming that the stirring is balanced by another mechanism.  This determines our two growth modes: \sumi\ in equilibrium growth $T_\mathrm{stir}$ is balanced by the friction time $T_\mathrm{fr}$, whereas \sumii\ in non-equilibrium growth $T_\mathrm{stir}$ is balanced by the growth timescale of the protoplanet, $T_\mathrm{grow}$. 

\subsubsection{Equilibrium growth: $T_\mathrm{stir} = T_\mathrm{fr}$.}
\label{sec:equil}
In this regime $T_\mathrm{stir} = T_\mathrm{fr} < T_\mathrm{gr}$.  In the typical case that both $T_\mathrm{stir}$ and $T_\mathrm{fr}$ depend on eccentricity, we use an iterative procedure to solve for $e_h$.  For example:
\begin{enumerate}
  \item[0.] Start with an initial value of $v_\mathrm{gas}$, \eg\ $v_\mathrm{gas}=v_\mathrm{hw}$ .
  \item Using $v_\mathrm{gas}$, obtain the friction time according to \eq{tdrag}. 
  \item Equate $T_\mathrm{stir}=T_\mathrm{fr}$ and solve for the eccentricity $e_h$ by means of the two segment fit presented in \fg{Tvs}.
  \item Determine a new $v_\mathrm{gas}$ using \eq{vgas} and the value of $e_h$ derived in the previous step.  
\end{enumerate}
We iterate the loop until convergence is achieved.  We then obtain $e_{h,k}$ and $T\sub{gr}{k}$ via \eq{Tgr}. Finally, we check for the validity of the equilibrium regime: of the three timescales involved, $T\sub{gr}{k}$ must be the longest.  If this condition is not satisfied, we turn to the nonequilibrium mode instead.

\subsubsection{Nonequilibrium growth: $T_\mathrm{stir} = T_\mathrm{gr}$.}
\label{sec:noneq}
In this regime $T_\mathrm{stir} = T_\mathrm{gr} < T_\mathrm{fr}$.  We numerically solve
\begin{equation}
  T_\mathrm{stir}(e_h) = T_\mathrm{gr}(e_h)
  \labelH{eq:ne-reg}
\end{equation}
for $e_h$.
We then obtain the (Hill) eccentricity consistent with the non-equilibrium regime.  For simplicity, when solving the above equation we assume that $T_\mathrm{gr}$ is independent of the friction time (\ie\ we ignore drag or atmosphere enhancement effects). 
This is justified, since the validity of the non-equilibrium solution is limited to the initial stages (low $T_\mathrm{gr}$) and big planetesimals (large $T_\mathrm{fr}$).

\subsubsection{Friction regime: $T_\mathrm{fr}=T_\mathrm{gr}$}
When both the equilibrium and the nonequilibrium regime are inapplicable, we obtain $e_h$ consistent with the friction regime: $T_\mathrm{fr} = T_\mathrm{gr} < T_\mathrm{stir}$. We found that the friction regime may become important after planetesimals had been excited to a large eccentricity by turbulent stirring.

\subsection{Collisional fragmentation}
\labelH{sec:fragmt}
\subsubsection{Material strength and the collision timescale}
The collision timescale for particles within the same component is given by $T\sub{col}{kk}^{-1} = n_k \sigma_{kk} \Delta v_{kk}$ ($k=$ P, F), where $n_k$ is the number density, $\sigma_{kk}$ the collision cross section, and $\Delta v_{kk}$ the relative velocity between two $k$-particles.  We relate $n_k$ to $\Sigma_k$ via
\begin{equation}
  n_k = \frac{\Sigma_k\left/ \frac{4\pi}{3}\rho_s s_k^3 \right.}{2h_k}
\end{equation}
with $h_k$ the scaleheight of the particle layer (\eqp{hp}).  For equal-size spheres, $\sigma_{kk} = 4\pi s_k^2$ and
\begin{equation}
  T\sub{col}{kk} = \frac{2\rho_s s_k h_k}{3\Sigma_k \Delta v_{kk}}.
  \labelH{eq:Tcol}
\end{equation}
For `heavy' particles like planetesimals the relative velocity and scaleheight are determined by the Keplerian elements $e_k$ and $i_k$. Then, $\Delta v_{kk} = e_kv_K \approx 2i_ka_0 \Omega = 2h_k \Omega$ and
\begin{equation}
  T\sub{col}{kk} = \frac{\rho_s s_k}{3\Sigma_k \Omega};\qquad (\tau_\mathrm{fr} \gg1).
  \labelH{eq:Tcolk}
\end{equation}

\begin{figure}[t]
  \plotone{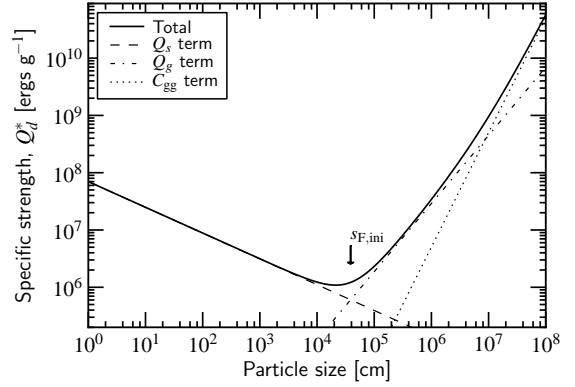}
  \caption{Specific strength of bodies as function of their size (\textit{solid} curve).  The material properties reflect those of ice.  The thin dashed, dashed-dotted, and dotted lines denote the three contributions out of which $Q_d^\ast$ is composed.  The arrow indicates the initial fragment size.}
  \labelH{fig:Qdstar}
\end{figure}
In the toy model we assume that the fragmentation rate is determined by the ratio of the specific collision energy and the specific strength of the material:
\begin{equation}
  q_k \equiv \left. \frac{1}{2}m_\mu (\Delta v_{kk})^2 \right/ (m_1+m_2)Q_d^\ast 
  = \frac{(\Delta v_{kk})^2}{8Q\sub{d}{k}^\ast}
  \labelH{eq:qi}
\end{equation}
where $m_1,m_2$ are the masses of the collision partners and $m_\mu$ the reduced mass.  For equal-mass particles $m_1=m_2=2m_\mu =m_k$. The material strength $Q_d^\ast$ is assumed to obey the following size dependence:
\begin{equation}
  Q_d^\ast(s) = Q_{0s} \left(\frac{s}{\mathrm{1\ cm}}\right)^{b_s} +Q_{0g} \rho_s \left(\frac{s}{\mathrm{1\ cm}}\right)^{b_g} +C_\mathrm{gg}v_\mathrm{esc}^2(s),
  \labelH{eq:Qd}
\end{equation}
where the three terms on the RHS respectively denote contributions from the strength regime, the gravity regime, and the gravitational potential.  The constant $C_\mathrm{gg}$ is fixed at $9$ \citep{StewartLeinhardt2009}. In \fg{Qdstar} we plot $Q_d^\ast$ as function of size for ice-like materials \citep{BenzAsphaug1999}.  The corresponding values are also listed in \Tb{constants}.

The fragmentation among planetesimals of size $s_P$ increases the abundance of fragments of size $s_F\ll s_P$.  The intermediate scales -- \ie\ the collisional cascade -- is not included in the toy model.  Erosive collisions between fragments and planetesimals can be included (\texttt{ER=1}), which results in an increase of the fragmentation rate.  We now provide expressions for the fragmentation rate $\dot\Sigma_{PF}$ due to mutual collisions among planetesimals and erosive collisions with fragments, respectively.

\subsubsection{Planetesimal-planetesimal collisions}
We adopt the prescription of \citet{KobayashiTanaka2010} for the excavated mass ($m_e$) in a fragmenting collision:
\begin{equation}
  m_e = \frac{\phi_P}{1+\phi_P}(m_1+m_2),
  \labelH{eq:me}
\end{equation}
where $\phi$ is defined as
\begin{equation}
  \labelH{eq:phi}
  \phi = \frac{\frac{1}{2}m_\mu (\Delta v)^2}{(m_1+m_2)Q_d},
\end{equation}
For equal-mass planetesimals, $\phi=\phi_P = (\Delta v_\mathrm{PP})^2/8Q_\mathrm{d,P} = q_P$.  Collisions take place on a planetesimal collision timescale $T_\mathrm{col,PP}$ (\eqp{Tcolk}), during which each body loses a fraction $q_P/(1+q_P)$ of its mass, resulting in a fragmentation rate of
\begin{equation}
  \frac{d\Sigma_P}{dt} = - \frac{q_P}{(1+q_P)} \frac{\Sigma_P}{T_\mathrm{col,PP}}.
  \labelH{eq:dsigPdt-P}
\end{equation}
In the limit of $q_P\gg1$ \eq{dsigPdt-P} reduces to $d\Sigma_P/dt = -\Sigma_P/T_\mathrm{col,PP}$ and the surface density in planetesimals is reduced on a collision timescale.  For $q_P\ll1$ the depletion timescale increases by a factor $q_P^{-1}$.

\subsubsection{Planetesimal-fragment collisions (erosion)}
\label{sec:eros}
For planetesimal-fragment collisions we can assume that the planetesimals determine the relative velocity ($\Delta v_\mathrm{PF} = \Delta v_\mathrm{PP}$) and cross section, $\sigma_\mathrm{PF}=\pi s_P^2$.  \Eq{phi} then reads
\begin{equation}
  \labelH{eq:phi-eros}
  \phi = \frac{m_F (\Delta v_\mathrm{PF})^2}{2m_P Q_\mathrm{d,P}} = \frac{4q_P m_F}{m_P} \ll 1.
\end{equation}
The relevant collision timescale is now the timescale on which a fragment collides with a planetesimal
\begin{equation}
  T_\mathrm{col,PF}^{-1} = n_\mathrm{eff} (\Delta v_\mathrm{PF}) \sigma_\mathrm{PF}. 
\end{equation}
The above equation must account for the fact that the two populations have different scaleheights: if $h_F \gg h_P$ (resulting, for example, from turbulent diffusion) the density is set by the scaleheight of the fragments; otherwise (if $h_F \ll h_P$) by that of the planetesimals.  Thus, $n_\mathrm{eff}$, the effective density of fragments for the collision, involves the largest scaleheight and we write $n_\mathrm{eff} = \Sigma_F/2h_\mathrm{max} m_F$ with $h_\mathrm{max} = \max(h_F, h_P)$:
\begin{equation}
  T_\mathrm{col,PF}^{-1} = \frac{\pi s_P^2 \Sigma_F}{m_F} \left( \frac{h_P}{h_\mathrm{max}} \right) \Omega,
\end{equation}
where we used $\Delta v_\mathrm{PP} = 2h_P \Omega$.  During the collision, the planetesimal loses a fraction $\phi$ (\eqp{phi-eros}) of its mass.  The fragmentation rate then reads
\begin{eqnarray}
  \labelH{eq:dsigPdt-E}
  \frac{d\Sigma_P}{dt} 
 &=&-\frac{\phi \Sigma_P}{T_\mathrm{col,PF}} 
  = -\frac{4 q_P m_F \Sigma_P \pi s_P^2 \Sigma_F \Omega}{m_Fm_P} \left( \frac{h_P}{h_\mathrm{max}} \right) \\
 &=&-\frac{q_P \Sigma_F}{T_\mathrm{col,PP}} \left( \frac{h_P}{h_\mathrm{max}} \right).
\end{eqnarray}
where $m_P= 4\pi \rho_P s_P^3/3$ and \eq{Tcolk} have been used.  A large erosion rate is obtained when $h_\mathrm{max}$ is given by the planetesimal layer (requiring $h_P \ll h_F$) and when $q_P\gg1$.  Under these conditions fragmentation by fragment-planetesimal collisions outcompetes fragmentation by planetesimal-planetesimal collisions. 

Adding the contribution from planetesimal-planetesimal collisions and fragment-planetesimal collisions, \eqs{dsigPdt-P}{dsigPdt-E}, gives a combined fragmentation rate of:
\begin{equation}
  \dot\Sigma_\mathrm{PF} 
  = -\frac{q_P}{1+q_P} \frac{\Sigma_P}{T_\mathrm{col,PP}} -q_P \frac{\Sigma_F}{T_\mathrm{col,PP}} \left( \frac{h_P}{h_\mathrm{eff}} \right).
  \labelH{eq:sigPdt-tot}
\end{equation}

\subsubsection{Change in the characteristic size of the planetesimals and fragments.}
\labelH{sec:s-change}
The planetesimal and fragment population are assigned a characteristic size ($s_F$ and $s_P$).  In the toy model it is possible to follow the coagulation of planetesimals, increasing $s_P$ with time.  If this feature is implemented (\texttt{PG=1}), planetesimals grow on a collision timescale (\eqp{Tcol}):
\begin{equation}
  \frac{ds_P}{dt} = \frac{s_P}{3T_\mathrm{col,PP}}\quad \texttt{(PG=1)},
  \labelH{eq:dsPdt}
\end{equation}
Thus, for planetesimal collisions we can have the somewhat counterintuitive situation that they grow (increasing $s_P$) while at the same time lose surface density due to fragmentation (\eqp{sigPdt-tot}).

Since bodies in the gravity regime become weaker when they are smaller, fragmentation of planetesimals triggers a collisional cascade.  In our toy model the fragment mass ($s_F$) represents the lower range of the cascade.  As the initial value for $s_F$ we choose the point where the strength and gravity contributions to $Q_d^\ast$ are equal (see \fg{Qdstar}).  Thus, mass liberated at $s=s_P$ flows instantaneously towards $s=s_F$.  Typical initial sizes for $s_F$ are of the order of 100 m (for basalt or icy planetesimals).  We only decrease $s_F$ when fragment-fragment collisions become disruptive.  The fragment size is then given by the implicit equation $q_F(s_F) = v_F^2/8Q_d^\ast(s_F) = 0.5$, where $v_F$ is the random motion among fragments and the 0.5 value follows from comparison against \citet{KobayashiEtal2010}.

\subsection{Treatment of solid's radial motions}
\labelH{sec:radial}
In protoplanetary disks, solids are expected to move radially.  Embryos may migrate due to type-I migration, planetesimals can diffuse or scatter due to gravitational encounters, and radial orbital decays due to drag forces.  In our context, radial drift of small particles (fragments) is especially important since it operates on short timescales.  To capture this effects, we will assign a characteristic drift timescale for the planetesimal and a fragment components, $T\sub{dr}{k}$.  In addition, the toy model can account for influx (accretion) from external regions ($\dot{M}_\mathrm{ext}\neq0$).  Nevertheless, since our model is local, it only crudely captures the nature of these radial motions; a multidimensional extension (see below) is needed for a self-consistent treatment.

Let us denote by $W_\mathrm{sim}$ the radial width of the local space around $a_0$ that is modeled by the toy model. The mass in solids, planetesimals ($k=P$) or fragments ($k=F$), is denoted $M\sub{W}{k}$ and changes according to
\begin{equation}
  \labelH{eq:dMdt-k}
  \frac{dM\sub{W}{k}}{dt} = 2\pi a \Sigma'_k \frac{dW_\mathrm{sim}}{dt} + \dot{M}\sub{ext}{k} + \dot{M}_\mathrm{int} -\frac{M\sub{W}{k}}{T\sub{dr}{k}},
\end{equation}
where the terms on the RHS of \eq{dMdt-k} denote, respectively: \sumi\ the increase in mass due to the growth of the feeding zone (the surface density outside $W_\mathrm{sim}$ may be different from $\Sigma$ and is denoted $\Sigma'$); \sumii\ the increase due to accretion from external regions;  \sumiii\ the change in mass within $W_\mathrm{sim}$ due to transfer of mass to a component other than $k$, \eg\ fragmentation (\eqp{sigPdt-tot}) and embryo growth; and \sumiv\ the mass loss from $W_\mathrm{sim}$ due to radial drift.  Since the surface density is $\Sigma_k = M\sub{tot}{k}/2\pi a W_\mathrm{sim}$ \eq{dMdt-k} translates into
\begin{equation}
  \labelH{eq:dSigkdt}
  \frac{d\Sigma_k}{dt} = \frac{1}{W_\mathrm{sim}}\frac{dW_\mathrm{sim}}{dt} (\Sigma'_k - \Sigma_k) + \frac{\dot{M}\sub{ext}{k}}{2\pi a W_\mathrm{sim}} + \dot{\Sigma}\sub{int}{k} -\frac{\Sigma_k}{T\sub{dr}{k}}.
\end{equation}
with
\begin{equation}
  \labelH{eq:Tdrift}
  T\sub{dr}{k} = \frac{W_\mathrm{sim}}{v\sub{dr}{k}};\qquad (\texttt{RD=1})
\end{equation}
When radial drift is not implemented (\texttt{RD=0}), one can formally write $T\sub{dr}{k}=\infty$.  \Eq{dSigkdt} very generally describes the evolution of $\Sigma_k$. But the length scale $W_\mathrm{sim}$ yet needs to be specified. Here, we consider two choices for $W_\mathrm{sim}$ reflecting the limits of a local or a global application of \eq{dSigkdt}.

\subsubsection{Local interpretation}
In a local model $W_\mathrm{sim}\ll a_0$ although $W_\mathrm{sim}$ should be chosen larger than the feeding/stirring zone of the embryo, \ie\ $W_\mathrm{sim}>\tilde{b}R_h$.  For such small $W_\mathrm{sim}$, the drift timescale $T\sub{dr}{k}$ will generally be short (especially for fragments). Boundary effects, that is the mass flow $\dot M\sub{ext}{k}$ coming into $W_\mathrm{sim}$, are important and must be quantified.  Ideally, $\dot{M}_\mathrm{ext}$ is obtained from the neighboring annulus as part of a multi-zone setup, $\dot{M}\sub{ext}{k} = 2\pi a' \Sigma' v'\sub{dr}{k}$, where primes denote quantities from the annulus exterior to $W_\mathrm{sim}$.  Although a multi-zone extension allows for a self-consistent treatment, it also considerably increases the overhead of the calculations. We will not implement it here, but may choose do so in an upcoming work.  

Generally, therefore, we must specify the accretion rate, $\dot{M}\sub{ext}{k}(t)$, as function of time. The picture here is of planet-formation going on in zone $W_\mathrm{sim}$, with solids exterior to $W_\mathrm{sim}$ steadily drifting in. In \se{ext-accr} we consider such a scenario as a simple application of the toy model in its local setting.

\subsubsection{Global interpretation}
In a global setting, we assume that the physical conditions are similar for scales on the order of the disk radius, $W_\mathrm{sim} \approx a_0$.  We no longer bother about mass in- or outflow due to growth of protoplanets or accretion from external regions (these are boundary effects).  The first two terms of \eq{dSigkdt} are put 0.  Specifically, we write
\begin{eqnarray}
  \labelH{eq:dSigkdt-ol}
  \frac{d\Sigma_k}{dt} &=& \dot{\Sigma}_\mathrm{int} -\frac{\Sigma_k}{T\sub{dr}{k}} \\
  T\sub{dr}{k} &=& \frac{W_\mathrm{sim}}{v\sub{dr}{k}} = C_\mathrm{drift} \frac{a_0}{v\sub{dr}{k}},
  \labelH{eq:vdr-ol}
\end{eqnarray}
where $C_\mathrm{drift}\approx1$ is a fudge factor.  We found that $C_\mathrm{drift}=0.5$ gives the best correspondence with multi-zone simulations models.

The global extrapolation is tailored towards the oligarchic growth phase, since we can assume that neighboring oligarchs are of similar mass (although this assumption will break down on scales $\sim$$a_0$).  Thus, when the switch $\texttt{OL}$ is set, \eqs{dSigkdt-ol}{vdr-ol} apply.  On the other hand, the local scenario (\eqps{dSigkdt}{Tdrift}) applies when we consider a \textit{single} protoplanet (\texttt{OL=0}).

\subsubsection{The surface density evolution timescale $T_\Sigma$}
\labelH{sec:mtr}
\Eq{sigPdt-tot} is one of the terms contributing to the internal mass flow $\dot\Sigma_\mathrm{int}$ in \eqs{dSigkdt}{dSigkdt-ol}: it re-arranges the mass among the three components, but conserves the total surface density in $W_\mathrm{sim}$.  Embryo growth also transfers mass: $\dot\Sigma_{k\mathrm{E}} = -C_\mathrm{acc}^{-1} \Sigma_E/T\sub{gr}{k}$.  Note the reduction by $C_\mathrm{acc}^{-1}$ as, when oligarchy is considered, a third of the embryo growth comes from embryo-embryo collisions (see \se{key-defs}).

Together, these two terms (\ie\ \eqp{sigPdt-tot} and $\dot\Sigma_{k\mathrm{E}}$) give the total internal mass flow $\Sigma\sub{int}{k}$.  Let us for simplicity denote the first two terms of \eq{dSigkdt} by $\dot\Sigma\sub{ext}{k}$, which represent the mass flow from exterior regions or from the expansion of the feeding zone. This term is therefore zero when \texttt{OL=1}.  Then, we can combine \eqs{dSigkdt}{dSigkdt-ol} into
\begin{equation}
  \dot\Sigma_k = \dot\Sigma\sub{int}{k} +\dot\Sigma\sub{ext}{k} -\frac{\Sigma_k}{T\sub{dr}{k}}.
\end{equation}
where $T\sub{dr}{k}$ is given either by \eq{Tdrift} or \ref{eq:vdr-ol}. The surface density then evolves on a timescale of
\begin{equation}
  \labelH{eq:TsigF}
  T^{-1}\sub{\Sigma}{k} = \frac{\dot\Sigma\sub{int}{k} +\dot\Sigma\sub{ext}{k}}{\Sigma_k} -\frac{1}{T\sub{dr}{k}}.
\end{equation}
Very often, when $T\sub{dr}{k}$ is short (\ie\ for small fragments), a semi-steady state is reached where $\Sigma_k \approx (\dot{\Sigma}\sub{int}{k} + \dot{\Sigma}\sub{ext}{k})T\sub{dr}{k}$.  In that case, $|T\sub{\Sigma}{k}| \gg T\sub{dr}{k}$ and it is computationally advantageous to evolve the evolution on timescales of $T\sub{\Sigma}{k}$ instead of the much shorter $T\sub{dr}{k}$.
  
\subsection{Summary of the algorithm}
\begin{deluxetable*}{lllllllllllp{8cm}}
  \tablecaption{\labelH{tab:sim-switch}Switch and parameter values}
  \tablehead{
  Section       &Figure  & \multicolumn{9}{l}{Switch value} & Parameters deviating from standard value or range\\
  &        & \texttt{AE} & \texttt{ER} & \texttt{FR}  & \texttt{ND}  &\texttt{OL} & \texttt{PG} & \texttt{RD} & \texttt{TD} & \texttt{TS}
    }
  \startdata
  \ref{sec:chamb-comp}  & \Fg{chamb-comp}   & x   & x   & x   & 0    & 1  & 0   & 0   & 0   & 0 & $\Sigma_\mathrm{ini} = 7\ \mathrm{g\ cm^{-2}}$  \\
  \ref{sec:koby-comp10} & \Fg{koby-comp}    & 0   & x   & x   & 0    & 1  & 1   & 1   & 0   & 0 & $a_0=3.2$ AU \\
  \ref{sec:koby-comp11} & \Fg{parstud}      & x   & 1   & 1   & 0    & 1  & 1   & 1   & 0   & 0 & $a_0=3-35$ AU;\quad $\Sigma_\mathrm{ini} = 1-10$ km;\quad $s_\mathrm{P,ini} =$ 1--100 km\\
  \ref{sec:switches}    & \Fg{jupiter}a     & 0   & 1   & 1   & 0    & 1  & 1   & 1   & x   & x & $\alpha_\mathrm{ss}=10^{-4}-0.1$ \\
  \ref{sec:switches}    & \Fg{jupiter}b     & x   & 1   & 1   & x    & 1  & 1   & 1   & x   & 0 & $Q_\mathrm{0s}=Q_\mathrm{0g}=0$ \\
  \ref{sec:switches}    & \Fg{jupiter}c     & x   & 1   & 1   & 0    & 1  & 1   & 1   & 0   & 0 & $v_\mathrm{hw}=2-54\ \mathrm{m\ s^{-1}}$ \\
  \ref{sec:ext-accr}    & \Fg{ext-accret}   & x   & 1   & 1   & x    & 0  & 0   & 1   & x   & 0 & $s_F = 10^{-1}-10^{4}$ cm;\quad $\Sigma_P=0$;\quad $M_\mathrm{ini}=10^{-2}\ M_\oplus$;\quad $\dot{M}_\mathrm{ext}>0$ 
  \enddata
  \tablecomments{Overview of the simulations in \ses{tests}{applic} listing the behavior of the switches: 0 indicates the switch is turned off; 1 that it is switched on; and `x' that its influence is examined.  See \Tb{switches} for the abbreviation and description of the switches and \Tb{constants} for the default parameter values. }
\end{deluxetable*}
\labelH{sec:implement}
In essence, the algorithm consist of a loop in which the key quantities (the embryo mass $M_E$, surface densities $\Sigma_k$, and characteristic sizes $s_k$) are advanced over a timestep of arbitrary length, until a final time ($T_\mathrm{end}$) is reached.  Within each iteration $i$ we calculate:  \sumi\ the eccentricity of the fragment and size distribution at $t=t_i$; \sumii\ the appropriate timestep $\Delta t_i$; and \sumiii\ the updated values for $M_E$, $\Sigma_k$, and $s_k$ at $t_{i+1} = t_i + \Delta t_i$:

\begin{enumerate}
  \item Given an embryo mass $M_E$, a particle size $s_k$ and surface density $\Sigma_k$, we calculate the growth mode -- equilibrium or non-equilibrium growth -- as outlined in \ses{dyn-state}{eh-sol}. This provides the embryo growth time $T_\mathrm{gr,k}$, the (Hill) eccentricity of the planetesimals and fragments, $e\sub{h}{k}$, and the friction times $T\sub{fr}{k}$. Using the calculated eccentricities, we obtain the fragmentation rates as outlined in \se{fragmt} and the timescales $T\sub{\Sigma}{k}$ on which the fragment and planetesimal populations evolve, \eq{TsigF}.
  \item The next step concerns the choice of the timestep $\Delta t_i$. It is computationally efficient to choose a $\Delta t_i$ as large as possible.  Yet, we must resolve the system on the shortest of the timescales characterizing the evolution of the system: 
  \begin{equation}
    \Delta t_i = \xi \min \left\{ T_\mathrm{gr,P}; T_\mathrm{gr,F}; T_\mathrm{\Sigma,P}; T_\mathrm{\Sigma,F}; T_\mathrm{col,PP} \right\},
    \label{eq:delta-i}
  \end{equation}
  where $\xi$ is a control parameter (we choose $\xi=0.1$ throughout).  Note that $T_\mathrm{dr,F}$ does not enter \eq{delta-i}, as it typically is very short.  Instead, we resolve the evolution of the surface density.  The planetesimal-planetesimal collision timescale $T_\mathrm{col,PP}$ is included to resolve planetesimal growth (\se{s-change}).
\item After the timestep is assigned, the quantities are advanced and $t_{i+1}=t_i+\Delta t_i$.  The embryo mass is increased by an amount $\Delta M_E = M_E \Delta t_i/T_\mathrm{gr}$ where $T_\mathrm{gr}$ is the total growth timescale due to accretion of planetesimals and fragments, $T_\mathrm{gr}^{-1} = T_\mathrm{gr,P}^{-1} +T_\mathrm{gr,F}^{-1}$. The planetesimal and fragment sizes are changed according to the prescription outlined in \se{s-change}. Finally, we update the surface densities of embryos, planetesimals, and fragments, using $\Delta t_i$ and $T\sub{\Sigma}{k}$.
\end{enumerate}

\begin{figure*}[t]
  \plottwo{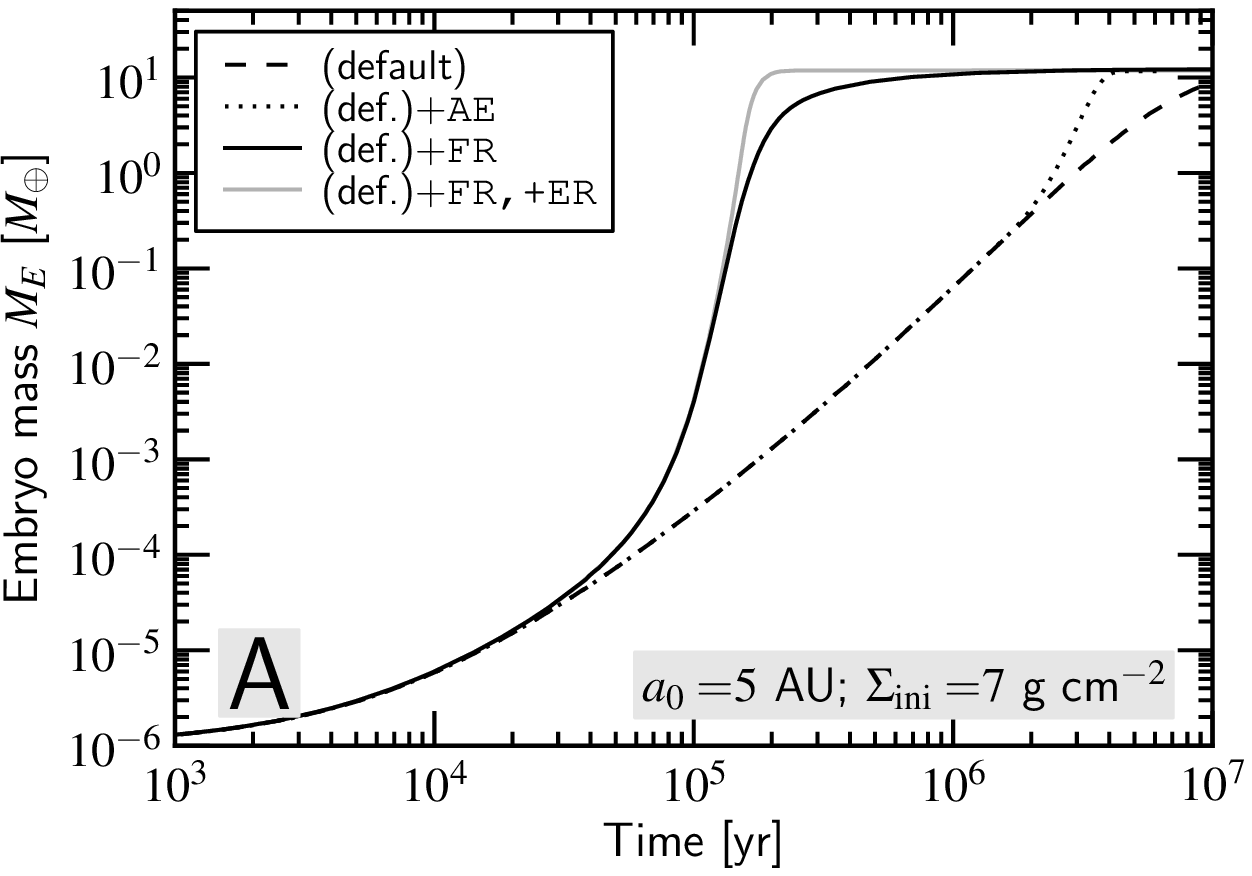}{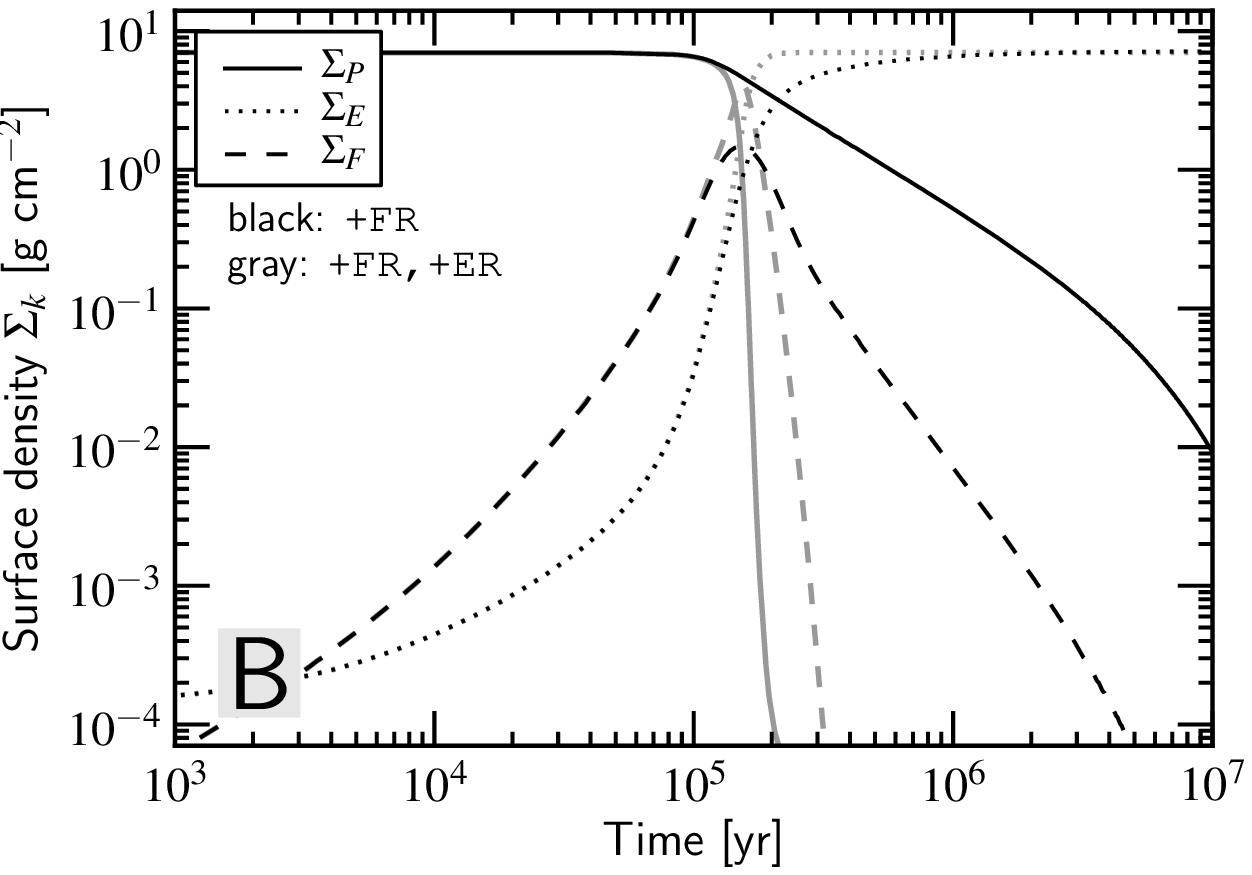}
  \caption{(\textit{left}) Sample runs at 5 AU, testing the influence of atmospheres and fragmentation on the growth of the embryo. The equilibrium solution for the dynamical state of the planetesimal population has been assumed and radial drift is switched off. The solid black line corresponds to the default run without fragmentation and atmosphere enhancement.  The \textit{dashed} curve corresponds to a run including planetesimal fragmentation but without accounting for erosive collisions (fragments colliding with planetesimals). The \textit{dotted} curve includes the erosive term. (\textit{right}) The surface densities in planetesimals, fragments, and embryos for the fragmentation run without erosion (\textit{black} curves) and including erosion (\textit{gray} curves).\\}
  \labelH{fig:chamb-comp}
\end{figure*}
\section{Comparison against previous studies}
\label{sec:tests}
We test the toy model against previous literature studies.  Throughout this and following section, we use the values listed in the second column of \Tb{constants} for the parameters of the toy model, unless specified otherwise.  In particularly, we adopt the \citet{BenzAsphaug1999} material parameters for ices (the $Q_{0s}$, $Q_{0g}$, $b_s$ and $b_g$ parameters).  Furthermore, we adopt the minimum mass surface nebula (MMSN) model \citep{Weidenschilling1977i,HayashiEtal1985} as our benchmark for the surface density and temperature structure of the disk:
\begin{eqnarray}
  \labelH{eq:mmsn1}
  \Sigma_g(a) &=& 4.0\times10^2\ \mathrm{g\ cm}^{-2}\ f_\mathrm{mmsn} \left( \frac{a}{\mathrm{5 AU}} \right)^{-1.5};\qquad  \\
  \labelH{eq:mmsn2}
  \Sigma_\mathrm{ini}(a) &=& f_\mathrm{gd}^{-1} \Sigma_g(a) ; \\
  \labelH{eq:mmsn3}
  T_g (a) &=& 125\ \mathrm{K}\ \left( \frac{a}{\mathrm{5 AU}} \right)^{-0.25} \left( \frac{L_\star}{L_\sun} \right)^{1/4};
\end{eqnarray}
where $f_\mathrm{gd}$ is the gas-solid mass ratio, assumed to equal 57 beyond the snowline (where ices contribute to the surface density), $f_\mathrm{mmsn}$ an enhancement factor for disks more massive than the MMSN, and $L_\star$ the stellar luminosity.  Note that \eqsto{mmsn1}{mmsn3} are merely a disguise for the local nature of the model, \ie\ the results depend on the local value of $\Sigma$, $\Omega$, $T_g$, only.  From the MMSN model we calculate the isothermal sound speed $c_s=\sqrt{k_BT_g/\mu}$, the disk scaleheight, $H_g = c_s/\Omega$, and the gas density at the midplane, $\rho_g = \Sigma_g/\sqrt{2\pi}H_g$.

The values of the switches and parameters deviating form their defaults are listed in \Tb{sim-switch} for each of the model runs.  Except indicated otherwise, we adopt an embryo mass of $M_\mathrm{ini,E}=10^{-6}\ M_\oplus$, assuming that by this stage the 2-component assumption characterizing oligarchy has materialized (see \ses{Intro}{overview}). Although in reality the start of the oligarchic phase depends on disk radius and initial planetesimal size \citep{IdaMakino1993,OrmelEtal2010}, we do not expect that changes in $M_\mathrm{ini,E}$ will matter greatly for the outcome of the results.

\begin{figure*}[t]
  \plottwo{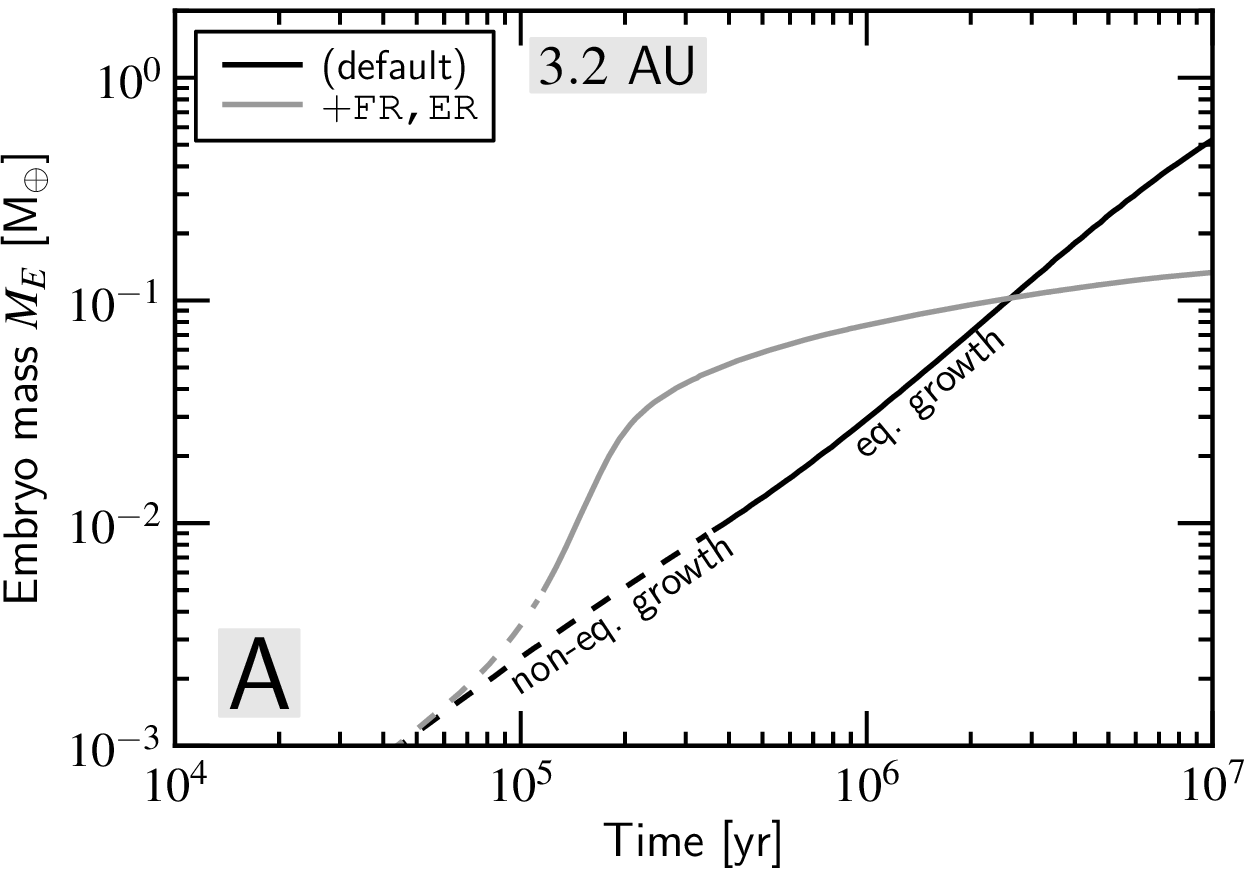}{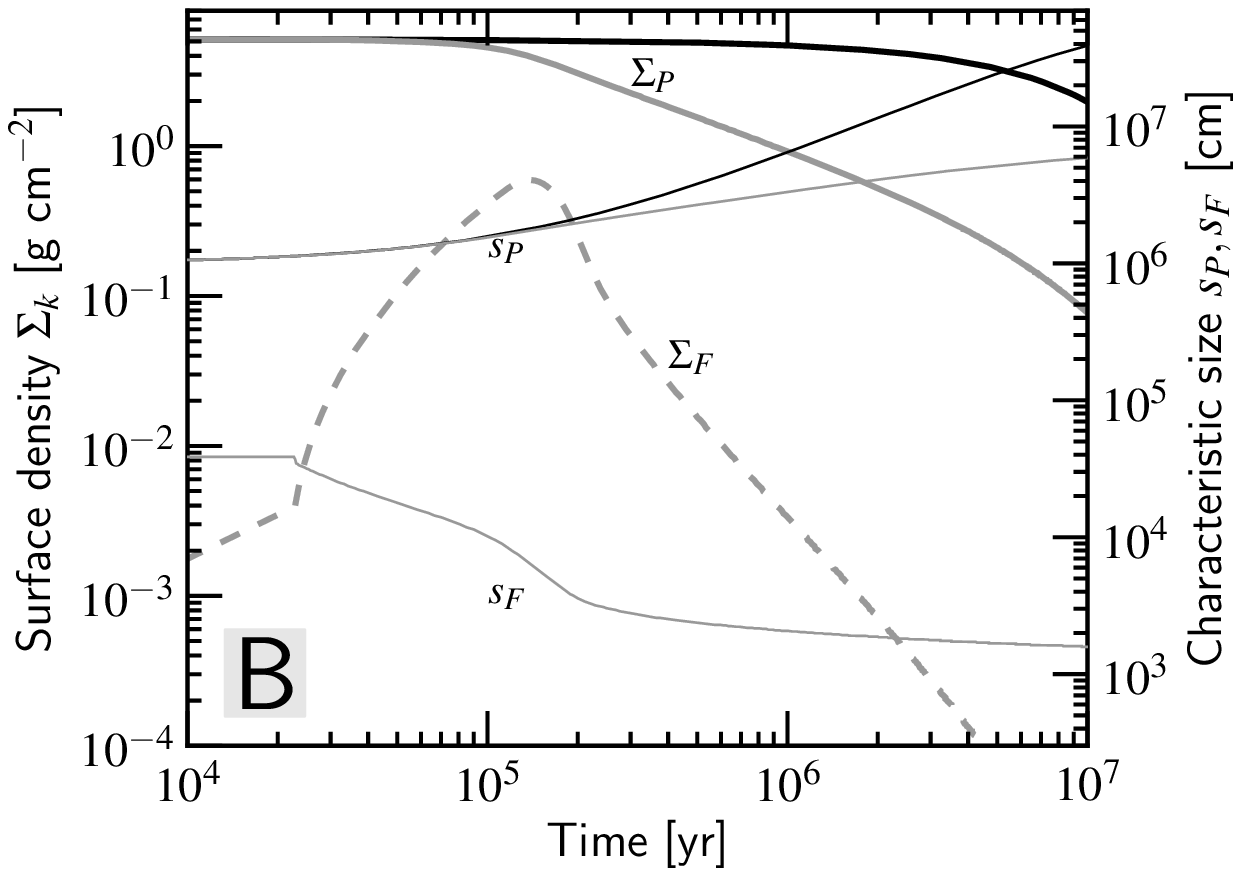}
  \caption{\textit{(left)} Mass of protoplanetary embryos \vs\ time for a run neglecting fragmentation (\textit{black} curve) and a run including fragmentation (\textit{gray} curve). The \textit{dashed} parts of the curves indicate that planetesimal growth follows the non-equilibrium solution (\se{noneq}); whereas the \textit{solid} parts correspond to the equilibrium solution (\se{equil}). \textit{(right)} The surface density in planetesimals ($\Sigma_P$, \textit{thick solid} curves), fragments ($\Sigma_F$, \textit{thick dashed} curve), and the characteristic sizes ($s_P$ and $s_F$; \textit{thin solid} curves) for the run without fragmentation (\textit{black} curves) and including fragmentation (\textit{gray} curves).\\}
  \labelH{fig:koby-comp}
\end{figure*}
\subsection{Comparison against \citet{Chambers2006}}
\labelH{sec:chamb-comp}
\citet{Chambers2006,Chambers2008} presents a model to follow the growth of embryos.  In \citet{Chambers2006} the basic model is introduced, whose elements are similar to ours: a 3-component model of planetesimals, fragments, and embryos.  Slowly, the complexity of the model is increased: \citet{Chambers2006} already includes the full calculation of the dynamical state, atmosphere enhancement of the capture radius, and radial drift. \citet{Chambers2008} includes a more detailed fragmentation model (with multiple size bins) and adds type-I migration, thereby including a spatial dimension (the disk radius) to the model.

We test our toy model against the 2006 (local) version of Chambers' model. This implies that we do not treat vertical diffusion of fragments, turbulent stirring, nebular drag effects, planetesimal growth, or erosive collisions: \texttt{TD=TS=ND=ER=PG=RD=0} (see \Tb{sim-switch}).  The model is conducted for a disk radius of 5 AU at a surface density of $7\ \mathrm{g\ cm^{-1}}$.  Furthermore, we only consider the equilibrium solution (the viscous stirring timescale balances friction timescale) to obtain the dynamical state of planetesimals.  Results are presented in \fg{chamb-comp} and should be compared to Figs.\ 6 and 8 of \citet{Chambers2006}.

With these assumptions our setups are nearly identical and we should expect to obtain very similar results.  The dashed curves labeled `(default)' in \fg{chamb-comp}a shows the embryo mass as function of time for the case without planetesimal fragmentation and without atmosphere enhancement.  The dotted curve shows the results when atmospheres are included (\texttt{AE=1}).  The run without atmospheres reaches the isolation mass $M_\mathrm{iso}$ only after $t=10^7$ yr, whereas the run that includes the increase of the collisional accretion rates due to atmospheres growth accelerates after $M_E\sim 0.1M_\oplus$, in agreement with \citep{Chambers2006}.  The isolation mass is the total solid mass of an embryo within an annulus of $\tilde{b}R_h$:
\begin{equation}
  M_\mathrm{iso} = \frac{\left(2\pi \tilde{b} \Sigma_\mathrm{ini} a_0^2\right)^{3/2}}{\sqrt{3M_\star}},
  \labelH{eq:Miso}
\end{equation}
which equals $M_\mathrm{iso}=11.8\ M_\oplus$ for the parameters in \fg{chamb-comp}.  When the embryo reaches the isolation mass, its surface density dominates, $\Sigma_E \gg \Sigma_F + \Sigma_P$, rendering the oligarchic approximations (\ie\ $e_E=0$) inapplicable (see \se{overview}).  But this only concerns the final doubling time in $M_E$.

In \citet{Chambers2006} fragments are assumed to lie in a dynamically cold ($e_F=0$) thin layer.  To mimic this effect, we fix the initial fragment size at 1 cm.  Gas drag will then ensure that the eccentricities are indeed negligible.  The accretion efficiency of these particles is very large (see \se{sd-reg}), and because fragments are \textit{not} assumed to drift (\texttt{RD=0}), fragmentation accelerates the growth significantly. The solid black curve in \fg{chamb-comp}a shows this behavior: after $t=10^5$ yr (roughly the collision timescale of the planetesimals) growth rates increase dramatically and the isolation mass is reached very quickly.  \Fg{chamb-comp}b, where the surface densities of the components are plotted further illustrates this point.  Although fragmentation is important, $\Sigma_F$ never becomes very large; fragments are very quickly accreted by the embryos.

The gray solid curve in \fg{chamb-comp}a and the gray curves in \fg{chamb-comp}b correspond to the outcome of a run that includes the erosion term (\ie\ the second term on the RHS of \eqp{sigPdt-tot}). This will be used in all of the following runs.  As can be seen from \fg{chamb-comp}b the surface density in planetesimals now declines precipitously after $t=10^5$ yr.  A positive feedback effect emerges: the higher the fragments' $\Sigma_F$, the higher $d\Sigma_F/dt$. In this laminar case ($\alpha_\mathrm{ss}=0$), the sandblasting of planetesimals occurs very fast, because the cm-size fragments are confined to a thin layer at the disk's midplane. 

\begin{figure*}
  \plotone{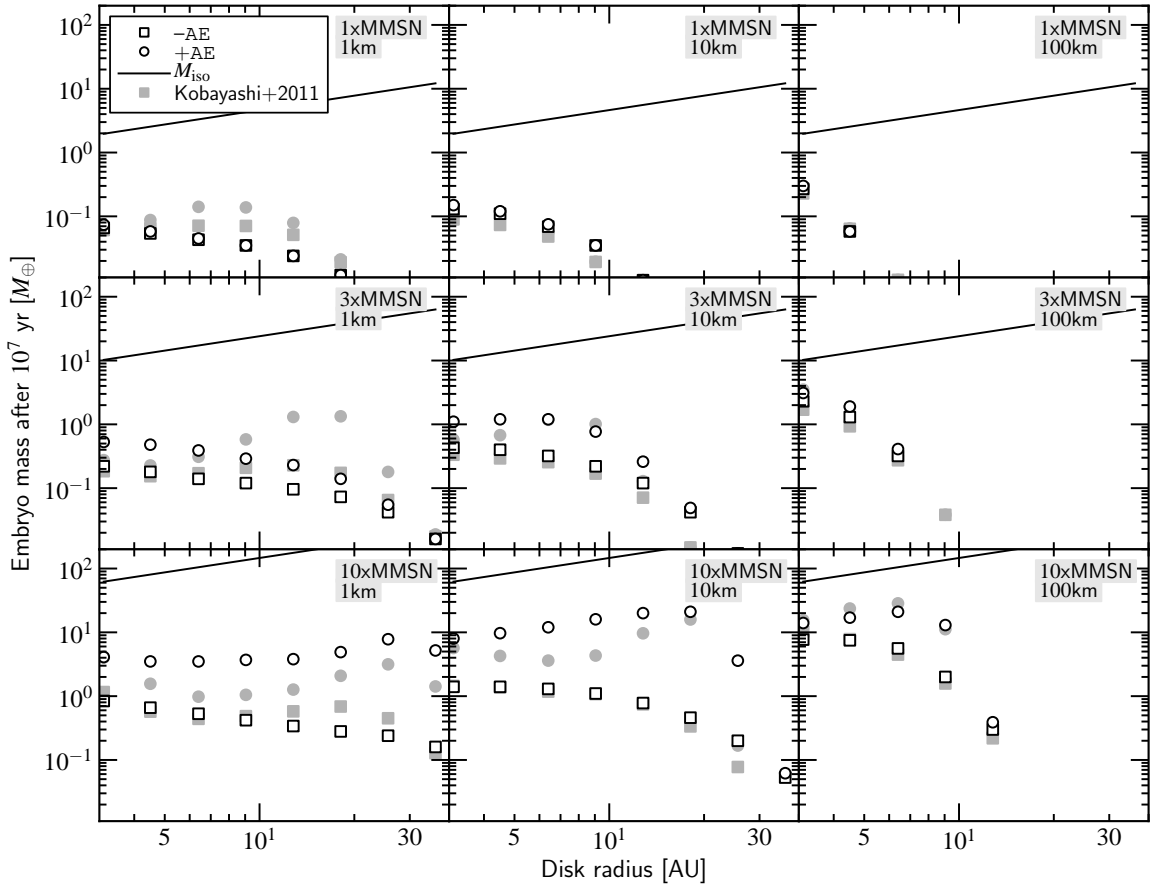}
  \caption{The embryos' masses after $10^7$ yr as function of disk radius for models that include fragmentation. \textit{Open black circles} correspond to models including atmosphere enhancement of the collisional cross section (\texttt{AE=1}), while \textit{open black squares} correspond to runs without atmospheres.  \textit{Note that each of these points corresponds to a separate run of our local model}.  The \textit{solid gray symbols} give the corresponding results of the multi-zone simulations of \citet{KobayashiEtal2011}. The solid line indicates the isolation mass $M_\mathrm{iso}$ (\eqp{Miso}). The nine panels each vary according to the disk mass with respect to the MMSN (\eqp{mmsn1}) and the initial size of the planetesimals $s_\mathrm{P,ini}$.\\}
  \labelH{fig:parstud}
\end{figure*}
\subsection{Comparison against \citet{KobayashiEtal2010}}
\label{sec:koby-comp10}
\citet{KobayashiEtal2010} presents a, multi-zone, multi-bin, model to follow the core formation.  They start from a monodisperse population of planetesimals and then compute the runaway growth and oligarchic growth stages. Their model is global: the total surface density of particles changes by radial drift motions from the outer disk to the inner disks.  Planetesimal fragmentation accelerates this inward flow of solids.  Indeed, \citet{KobayashiEtal2010} found that fragmentation was generally not conducive to growth, a somewhat different conclusion from \citet{Chambers2006,Chambers2008}.  \citet{KobayashiEtal2010} only considers laminar disks, and also ignore nebular drag effects.  Thus, the \texttt{TD, TS} and \texttt{ND} switches are all turned off.  The \texttt{PG} switch is turned on, whereas the \texttt{FR} switch can be turned on or off, depending on the model test. \citet{KobayashiEtal2011} include the effect of atmospheres.

We focus on the standard model of \citet{KobayashiEtal2010}, where the MMSN is assumed. \Fg{koby-comp}a shows the evolution of the embryo mass at a disk radius of 3.2 AU.  This plot should be compared to Fig.\ 7 of \citet{KobayashiEtal2010}.  The initial, dashed part of the curves indicate that planetesimals follow the non-equilibrium solution, whereas solid curves indicates the equilibrium solution (see \se{eh-sol}).  The black curve is a run without fragmentation; the gray curve includes the effects of planetesimal fragmentation and erosion.  

The growth-only run (black curve) follows a similar pattern as the corresponding curve of \citet{KobayashiEtal2010}.  After $3\times10^5$ yr the growth mode switches from non-equilibrium growth to equilibrium growth. An increase in the growth rate can then be expected since for the equilibrium regime gravitational focusing factors no longer decrease.  However, no steep increase is seen due to two opposing effects.  First, \textit{planetesimal growth} increases the mass of the planetesimals, which renders gas friction less effective and results in a smaller focusing factor (or larger $e_h$). Second, towards the end of the simulation, the planetesimal mass reservoir becomes empty.  \Fg{koby-comp}b illustrates these effects: $s_P$ (thin solid, black line) increases and $\Sigma_P$ (thick solid, black line) decreases towards the end of the simulation.

A somewhat more complex behavior can be seen when fragmentation is switched on (\texttt{FR=1}; gray curves in \fg{koby-comp}a,b). On timescales of $\sim$$T_\mathrm{col,PP}$ planetesimals collide and fragment.  The size of the fragments, determined from the criterion outlined in \se{s-change}, is initially $\sim$100 m and further decreases as fragments collide among themselves.  When their surface density is large, accretion of fragments can become very effective since their eccentricities are low.  The surface density in fragments peaks towards $t=10^5\ \mathrm{yr}$ at $\approx$10\%\ of the initial surface density in fragments. At this point the accretion rate is substantial.  However, the fragmentation and decreasing $s_F$ cause a stronger orbital decay as $\tau_\mathrm{fr}=\Omega T_\mathrm{fr}$ decreases towards unity (see \eqp{vr}).  As a result, $\Sigma_F$ quickly decreases.  In the end, then, the combined effect of fragmentation and radial drift stalls the growth as it depletes the solid mass reservoir.

\subsection{Including Atmospheres (\texttt{AE=1})}
\label{sec:koby-comp11}
We next run the toy model at several disk radii, pretending, to mimic the setting of a global model.  We will test the effects of the initial size of planetesimals ($s_\mathrm{P,ini}$), the inclusion of atmospheres (the \texttt{AE} switch), and the disk mass (which translates in a larger gas density $\rho_g$ and initial solid surface density $\Sigma_\mathrm{ini}$).  In total 144 models are run.  In \fg{parstud} we plot the final mass (after $10^7$ yr) of the runs. Black open squares corresponds to runs without atmosphere enhancement and black open circles to runs with \texttt{AE=1}.  These simulations should be compared to Figs.\ 3--5 of \citet{KobayashiEtal2011}, which data are shown in \fg{parstud} by solid gray symbols.

\begin{figure*}[t]
  \plotone{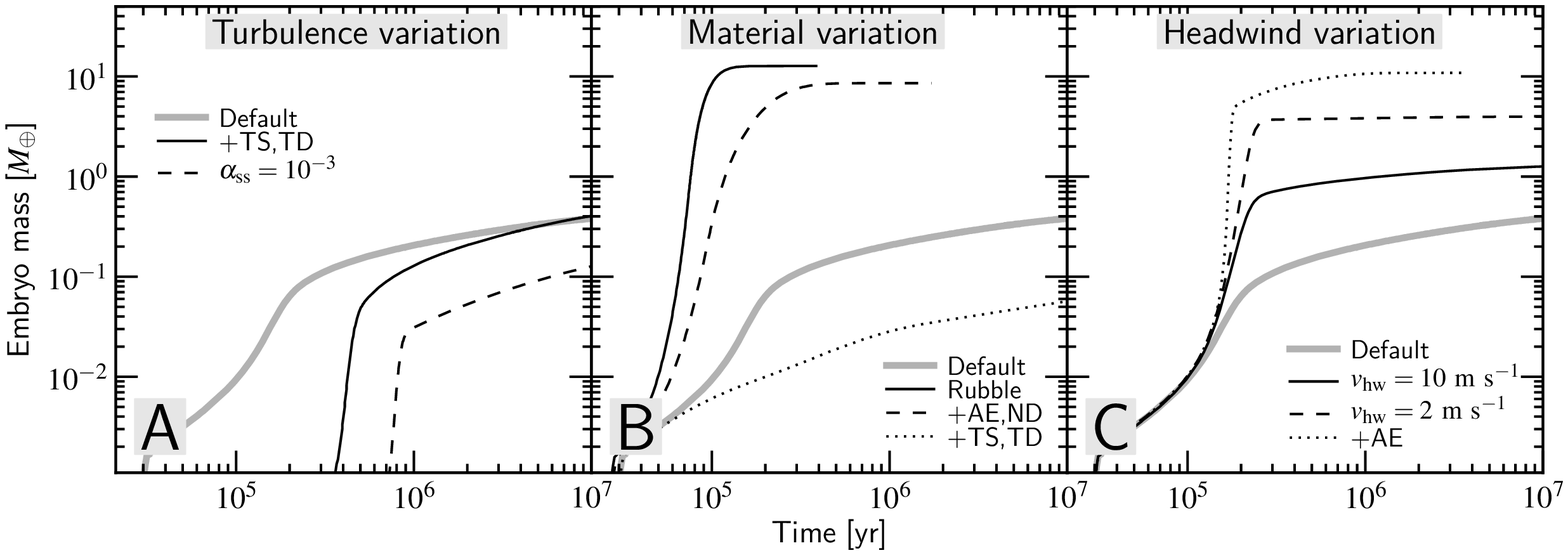}
  \caption{Sensitivity of embryo growth to the gas and material properties and \texttt{ND,TS,TD} switches. In each panel the thick gray line corresponds to the 5AU, 3xMMSN, 1 km run from \fg{parstud} with \texttt{ER=FR=OL=PG=RD=1} and \texttt{AE=ND=TS=TD=0}.  In \textit{(a)} we switch on turbulence by activating the \texttt{TS} and \texttt{TD} switches, and examine the sensitivity of the results to the turbulence-$\alpha_\mathrm{ss}$ parameter, whose standard value is $\alpha_\mathrm{ss}=10^{-4}$.  In \textit{(b)} we put $Q_{0s}=Q_{0g}=0$, indicative of a rubble-pile nature of the planetesimals without any internal strength. Collisions produce a copious amount of mm-size fragments, whose accretion behavior is very sensitive to the \texttt{AE}, \texttt{ND}, and \texttt{TD} switches.  In \textit{(c)} we reduce the disk headwind $v_\mathrm{hw}$ from its standard value of $54\ \mathrm{m\ s}^{-1}$.}
  \labelH{fig:jupiter}
\end{figure*}
Although the general correspondence is good, occasionally the predictions of the toy model are seen to deviate from the \citet{KobayashiEtal2010,KobayashiEtal2011} simulations.  The embryo masses for the 3xMMSN, 1 km runs at 10-20 AU in the toy model, for example, are lower than those of \citet{KobayashiEtal2011}, but the discrepancy especially becomes severe \textit{after} AE is switched on.  This is understandable, since, as we saw in \se{chamb-comp}, growth can accelerate dramatically when AE becomes important. The large discrepancy therefore merely reflects the sensitivity to the onset of this effect -- rather than indicating a fundamental flaw in the model. 

A more systematic difference is that for the toy model embryo masses show a stronger dependence on the disk mas (enhancement over MMSN).  Whereas the final embryo masses of the toy model mostly lag those of \citet{KobayashiEtal2011} at 1xMMSN, the toy model has overtaken the \citet{KobayashiEtal2011} simulations for the 10xMMSN runs, especially when AE is switched on.  This discrepancy may partly reflect our simplistic treatment of the atmosphere models (see \app{atmos}), where we assume that the opacity stay constant (and low) -- assumptions that may break down for massive atmospheres.  More fundamentally, strong drift motions of solids causes the breakdown of the oligarchy approximation: \ie\ that most of the mass is in planetesimals and fragments, rather than embryos.  For example, in the 10xMMSN,1 km run planetesimal fragmentation is very efficient and the embryos will dominate the surface density sooner rather than later.  In the \citet{KobayashiEtal2011} simulations the embryos then excite themselves to large eccentricities,\footnote{The simulations conducted by \citet{KobayashiEtal2011} did not include eccentricity damping due to gravitational torques with the gas disk \citep{TanakaEtal2002}.} which frustrates their growth.  In the toy model, however, $e_E=0$ is enforced and embryos always merge while keeping a separation of $\tilde{b} R_h$.

Altogether, the match of the toy model to the more sophisticated simulations by \citet{KobayashiEtal2011} is very satisfactory.  The general trends are similar, the final masses agree within a factor of 3 for most runs, and large discrepancies are mainly caused by the boost embryos experience due to crossing a certain mass threshold for the atmosphere effects kick in.  Our results reconfirm the main conclusion from \citet{KobayashiEtal2011}: that the planetesimal fragmentation-radial drift tandem makes it hard to grow large cores, even when atmosphere treatment is accounted for.  A great advantage of the toy model is that it is very fast: whereas it takes the \citet{KobayashiEtal2011} simulations 5--10 CPU hours on a modern PC to complete a single disk model, the eight corresponding runs of the toy model are finished in $\approx$10 seconds.

\section{Applications}
\labelH{sec:applic}
Due to its flexibility, there are countless imaginable applications that can be envisioned with the toy model.  A full investigation is beyond the scope of this paper.  Here, we illustrate the toy model by two examples: \sumi\ a modest sensitivity study to investigate in what way the results of the previous section depend on adopted gas and material properties; and \sumii\ the calculation of the accretion efficiency of fragments, drifting past a single embryo.  In both examples, we focus primarily on the behavior of small particles, as determined by the \texttt{AE}, \texttt{ND}, and \texttt{TD} switches.

\subsection{Sensitivity to gas and material properties}
\labelH{sec:switches}
Here we take a particular run of \fg{parstud} -- 5AU, 3xMMSN, 10 km, no atmosphere enhancement -- and explore how its results change if we change the adopted gas and material properties: the level of turbulence in the disk (\fg{jupiter}a); the material properties (\fg{jupiter}b) and the disk headwind parameter $v_\mathrm{hw}$ (\fg{jupiter}c).  In each panel the reference model of \fg{parstud} is indicated by the thick gray line.

In the toy model, turbulence is controlled by the dimensionless $\alpha_\mathrm{ss}$ parameter and is activated by setting one of the \texttt{TS} or \texttt{TD} switches.  Turbulent stirring (TS) due to density inhomogeneities in the gas may outweigh over viscous stirring when $\alpha_\mathrm{ss}$ is large, or when embryos are small (see \app{densflucstir}). Turbulent diffusion (TD) causes the particle layer to puff up, thereby decreasing the local densities of particles near the midplane.  \Fg{jupiter} shows that for a turbulence parameter of $\alpha_\mathrm{ss}=10^{-4}$ (solid curve) growth is initially very slow.  The slow growth is due to turbulent stirring, which excites the planetesimals to large, but constant eccentricities.  Over time, the Hill eccentricity $e_h$ decreases as the embryo grows; growth speeds up after a slow start and viscous stirring takes over eventually.  This `delayed' onset of (runaway) growth has been seen before \citep{OrmelEtal2010i,GresselEtal2011} and depends on the severity of the stirring, here parametrized by $\alpha_\mathrm{ss}$. 

In addition, large $\alpha_\mathrm{ss}$ will cause planetesimals to fragment `prematurely', \ie\ when the embryo is still small.  Fragments drift away before the embryo has an opportunity to accrete them, resulting in much smaller embryos.  This effect is reflected in the $\alpha_\mathrm{ss}=10^{-3}$ (dashed curve) and the $\alpha_\mathrm{ss}=10^{-2}$ (where embryo growth does not take off) runs.  There is a crucial difference between fragmentation triggered by turbulent stirring and viscous stirring.  When viscous stirring is important, the embryo is by definition massive and it will always sweep up some fraction of the solids.  But turbulent stirring is insensitive to the embryo mass, which renders it potentially much more harmful to embryo growth.

The size of the fragments and their production rate are largely determined by the material properties.  In \fg{jupiter}b we consider the case that the planetesimals are rubble piles by putting $Q_\mathrm{0s}=Q_\mathrm{0g}=0$, leaving the gravitational binding energy as the sole source of strength (\eqp{Qd}; \fg{Qdstar}).  Furthermore, the fragment size is fixed at 1 mm at all times. This setup is very conducive to growth, because the radial drift for mm-size particles is not so effective as it is for m-size particles.  As a result, embryos in the standard model (solid line) accrete these particles \textit{before} they drift away; and embryos grow large on a very short timescale ($\sim$$10^4$ yr!).  Nebular drag effects (\texttt{ND=1}) cause the fragments to be dragged along with the gas, rendering it somewhat more difficult to accrete them, especially when the protoplanet is small (dashed curve).  However, when the gas is only slightly turbulent ($\alpha_\mathrm{ss}=10^{-4}$; dotted curve) accretion virtually shuts down since the small particles are now distributed over a large scaleheight: embryo growth stalls at masses $<$$0.1\ M_\oplus$.  When small particles contribute significantly to the mass in solids, core formation will be very sensitive to the state of the turbulence.

Finally, we vary the value of the disk headwind $v_\mathrm{hw}$.  This parameter primarily determines the drift timescale, and hence the residence time of particles near the embryo.  \Fg{jupiter}c shows that a reduction by a factor of 5 (from the default value of $v_\mathrm{hw}=54\ \mathrm{m\ s}^{-1}$ to $10\ \mathrm{m\ s^{-1}}$) results in embryos that are a factor of three more massive.  Reducing the headwind parameter even further reinforces these findings. The 100 m-size `fragments' stay longer in the accretion zone, which enhances the likelihood that the embryo can capture them.  A larger capture radius due to atmospheres (\texttt{AE=1}) exacerbates these effects (dotted curve in \fg{jupiter}c).  Although in \fg{jupiter}c fragments are typically $\sim$10 m in size, the reduction of $v_\mathrm{hw}$ favors embryo growth for all fragment sizes.

The value of the disk headwind $v_\mathrm{hw}$, or, equivalently, the pressure gradient (\eqp{eta}), can be affected in several ways:
\begin{enumerate}
  \item Variations of the global pressure profile, \ie\ of the surface density and temperature index. From \eq{eta} one can show that $v_\mathrm{hw}$ depends on the exponents of the temperature and surface density of the gas.  Thus, deviations from the MMSN value $p=-3/2$ (as in $\Sigma \propto a^p$) will be directly reflected in $v_\mathrm{hw}$.  The same holds for a different temperature structure, \eg\ in optically thick disks. 
  \item Local changes in $\partial P/\partial a$.  More potent changes arise locally, if we account for variations in the accretion rate of the gas, triggered by dead zones \citep{Gammie1996}.  The interfaces between MRI dead and active zones in particular are promising candidates in this respect.  At these interfaces, the kinematic viscosity $\nu_T$ of the turbulence changes abruptly. To maintain a constant gas accretion rate ($\dot{M}_g \propto \nu_T \Sigma_g$) a decrease with disk radius in $\nu_T$ is accompanied by an increase in $\Sigma_g$ and a corresponding increase in the \textit{local} pressure gradient \citep{KretkeLin2007}.  Possibly, this even reverses the sign of the pressure gradient, such that drifting particles will accumulate in high pressure maxima.
  \item Collective effects. When dust particles are sufficiently concentrated they will act collectively and drag the gas with them in the direction of Keplerian velocities, in effect reducing the headwind a dust particle feels.  Here, we have not accounted for collective effect in our expressions for the radial drift (\eqps{vr}{vr2}), but this is straightforward to implement \citep{TanakaEtal2005,EstradaCuzzi2008,BaiStone2010i}.  Particle-hydrodynamic simulations often display strong collective effects (\eg\ \citealt{JohansenEtal2009}).
\end{enumerate}

\subsection{Efficiency of accreting inward-drifting particles by a single embryo}
\labelH{sec:ext-accr}
\begin{figure}[t]
  \plotone{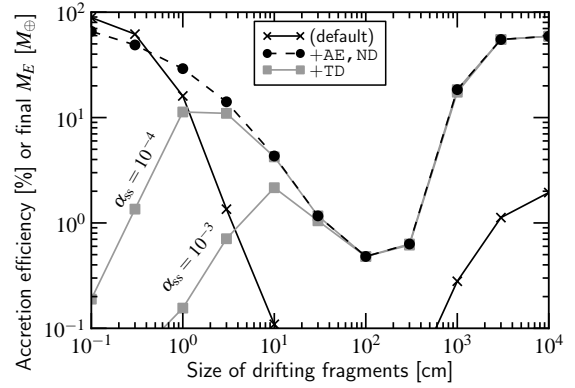}
  \caption{Accretion efficiency of inward-drifting particles by a single embryo. Each symbol corresponds to a run where a total mass of $100\ M_\oplus$ in fragments of size $s_F$ ($x$-axis) drifts past an embryo over a timescale of $\sim$10$^6$ yr. The fragment size does not evolve. Curves correspond to choices of the \texttt{AE},  \texttt{ND}, and \texttt{TD} switches: \texttt{AE=ND=TD=0} (\textit{black solid} line), \texttt{AE=ND=1} (\textit{dashed solid} line), \texttt{AE=ND=TD=1} (\textit{gray} lines with $\alpha_\mathrm{ss}=10^{-3}$ and $\alpha_\mathrm{ss}=10^{-3}$).\\}
  \labelH{fig:ext-accret}
\end{figure}
State of the art particle-hydrodynamical simulations show that planetesimals may form big \citep{JohansenEtal2009,CuzziEtal2010}, perhaps with radii up to $10^3$ km, out of a `sea' of small particles -- an idea that also finds support in the asteroidal record (\citealt{BottkeEtal2005,MorbidelliEtal2009}; but see \citealt{Weidenschilling2011} for an alternative interpretation). In this context, it is very relevant to study the interaction between embryos  and particles directly.  \citet{XieEtal2010} studied the sweeping-up of small particles by km-size planetesimals, and showed that the planetesimals acquired mass pretty fast, even though gravitational focusing was omitted.  Here, we extend their study to the regime of embryos.

We consider a scenario where particles from the outer disk drift past a single protoplanet situated at 5 AU.  For simplicity, we do not treat a planetesimal component ($\Sigma_P=0$); the planetesimal-related switches (\texttt{FR, ER, PG} and \texttt{TS}) are therefore irrelevant. The solids that drift inwards are all small particles of fixed size $s_F$. As we consider a single protoplanet the \texttt{OL} switch is off and the surface density evolution is described by \eq{dSigkdt}. The accretion rate $\dot{M}_\mathrm{ext}$ is given by
\begin{equation}
  \dot{M}_\mathrm{ext}(t) = \frac{M_\mathrm{tot}}{t_c} \exp[-t/t_c],
  \label{eq:Mext}
\end{equation}
where $M_\mathrm{tot}$ is the total, time-integrated, amount of solids that drifts by and $t_c$ the characteristic timescale.  Here, we choose $M_\mathrm{tot}=100\ M_\oplus$ and $t_c = 10^6\ \mathrm{yr}$.  It must be said that these parameters and \eq{Mext} are completely ad-hoc and purely chosen for illustrative convenience. We refer to \citet{YoudinChiang2004} and Birnstiel et al.\ (subm.) for physically-motivated prescriptions.  The initial mass of the embryo is fixed at $M_\mathrm{ini,E}=0.01\ M_\oplus$.  

The embryo mass after $t\gg t_\mathrm{ch}$ provides us with the accretion efficiency of the embryo.  This quantity is plotted in \fg{ext-accret} as function of fragment size.  Each symbol in this plot denotes a different run with the symbols connected by lines to guide the eye.  The default suite of models (diamonds) represents the case without accounting for drag effects or turbulent diffusion (\texttt{AE}=\texttt{ND}=\texttt{TD}=0).  One ontices that big boulders of 10--100 m in size have a $\approx$1\%\ (time-averaged) probability of being accreted.  Somewhat smaller, m-size particles fare much worse, though.  They are optimally coupled to the gas ($\tau_\mathrm{fr}=1$) and speed by at the maximum drift velocity of $v_\mathrm{dr}\approx v_\mathrm{hw}=54\ \mathrm{m\ s}^{-1}$ (see \eqp{vr}). Smaller particles, however, reside longer in the feeding zone of the planet and have a much higher probability of getting accreted -- for mm-size particles it becomes almost 100\%.

When we allow for drag-effects (\ie\ an atmosphere-enhanced capture radius, and nebular drag; dashed curve in \fg{ext-accret}), the accretion rate increases across the spectrum of sizes.  Heavier fragments now reach accretion probabilities of several tens of percents, sufficient to form large cores.  Accretion efficiencies also remain large for the smallest sizes, due to their long residence times in combination with their low scaleheights.  Turbulence, which will stir up these small particles, will therefore tend to negate these effects, see the gray curves in \fg{ext-accret}. For a (modest) $\alpha_\mathrm{ss}$ value of $10^{-4}$ it already shows much lower accretion efficiency for mm and cm-size particles; for $\alpha_\mathrm{ss}=10^{-3}$ the accretion efficiency is even lower.  These finding reflects those of \fg{jupiter}b. 

In conclusion, \fg{ext-accret} shows that the accretion efficiency of fragments can become sufficiently large to build sizable cores ($\approx$10 $M_\oplus$), although for smaller particles this depends heavily on the turbulence levels. 
MRI dead zones may therefore favor core growth, although recent studies indicate that particles are nonetheless very diffusive \citep{TurnerEtal2010,OkuzumiHirose2011}.  In addition, the results will depend on the headwind velocity, \ie\ the (local) pressure gradient of the gas.  \Fg{ext-accret} corresponds to $v_\mathrm{hw}=54\ \mathrm{m\ s^{-1}}$; but, as we saw from \fg{jupiter}c, a lower headwind will enhance the accretion rate.  The \citet{JohansenEtal2011} simulations, for examples, show that m-size boulders concentrate heavily.  These `particle bands' are dense enough to drag the gas along, such that the headwind vanishes.  Collective effect will mitigate low efficiencies seen in \fg{ext-accret} for $\tau_\mathrm{fr}=1$ particles.

\section{Summary and Outlook}
\label{sec:summ}
We have presented a new model to follow the protoplanet growth stage in planet formation.  The model is a toy model: it assumes that the mass distribution can be approximated by three components (embryos, planetesimals, and fragments) and does not involve a radial dimension. To obtain the equations that govern the evolution, we have employed simple physically-based recipes.  Yet the toy model is intended to be general: a wide array of physical relevant processes can be included at the discretion of the user.  In particular, we have focused on a more accurate treatment of fragments, for which coupling to gas is important.  Due to our modular setup (the switches) the toy model is intended to expose the key mechanisms that drive protoplanet growth. 

One should be aware of the limitations, as we have outlined in \se{overview}.  The (at most) three component approximations entails that the toy model cannot model the runaway growth phase (the phase preceding oligarchy) and does a bad job in modeling the fragmentation cascade.  The oligarchic assumptions will also break down when the mass in embryos ($\Sigma_E$) starts to exceed that of planetesimals and fragments -- a situation that may be reached sooner rather than later in cases where drift is important.  Furthermore, as of now, type I migration, or planetesimal driven migration \citep[][Ormel \& Ida in prep]{KirshEtal2009,CapobiancoEtal2011} are not included either. Our model is in it strictest sense only locally valid.

We have shown in \se{tests} that our results are in good agreement with those of \citet{KobayashiEtal2010,KobayashiEtal2011}.  These calculations indicate that planetesimal fragmentation (a natural outcome of oligarchic growth) is quite harmful to protoplanet growth, due to their strong orbital decay.  This conclusion contrasts with many earlier works, which suggest that small planetesimals, or fragments, are very conducive for growth \citep{Rafikov2004,GoldreichEtal2004,Chambers2006,KenyonBromley2009,GuileraEtal2011}.  However, \citet{Chambers2008} already warned that fragmentation acts like a double-edged sword -- increasing collision rates but at the same time depleting the disk of solids due to orbital decay -- whereas \citet{KobayashiEtal2011} finds that fragmentation is detrimental to embryo growth.  For this reason, they prefer bigger planetesimals since these are stronger and less collisional -- both effects reduce the fragmentation rate.  But the disadvantage of big planetesimals is that they are damped poorly and growth is relatively slow.

But it may still be premature to provide a final answer to the question whether small planetesimals, or small fragments, promote or hinder (fast) growth.  Very small particles have the additional handicap that turbulence stirs them up as we saw in \fg{jupiter}b and \fg{ext-accret}.  Few works concerning protoplanet growth have accounted for turbulent stirring (\citealt{BromleyKenyon2011i} are a recent exception), which is somewhat peculiar since $\alpha_\mathrm{ss}$ is a well known -- and critical -- parameter for the pre-planetesimal growth state \citep[\eg][]{BirnstielEtal2010i,ZsomEtal2011}.  Indeed, from an observation perspective mm/cm particles dominate (the mass) of protoplanetary disks and are the likely precursors of the first generation of big planetesimals c.q.\ protoplanet seeds \citep{JohansenEtal2007,JohansenEtal2009,CuzziEtal2010}.  Therefore, we better investigate the accretion behavior of small particles (\citealt{JohansenLacerda2010,OrmelKlahr2010}, \fg{ext-accret}). Here we reemphasize that our treatment of atmosphere enhancements (\texttt{AE}) and nebular drag effects (\texttt{ND}) is somewhat `ad-hoc' (see our comments towards the end of \se{atmos-enh}).

In a follow-up study we intend to return to the question regarding the efficiency of accretion of (big) planetesimal \vs\ (small) fragment.  The result of this work offers some tentative clues, though.  First, the efficiency scales with the magnitude of the headwind $v_\mathrm{hw}$ -- the deviation of the orbital velocity of the gas from Keplerian.  It seems very hard to grow big protoplanets ($\sim$10 $M_\oplus$) out of small particles when $v_\mathrm{hw}$ is not reduced or with some form of replenishment of solids.  Interestingly, a similar conclusion may hold for planetesimal formation itself \citep{CuzziEtal2010,BaiStone2010ii}. 

A second important point is that the accretion efficiency is strongly size dependent (\fg{ext-accret}). (Sub)km-size planetesimals accrete well, but are prone to collisional fragmentation.  The resulting m-size fragments fare much worse though in terms of accretion efficiency,  despite the fact that they have a large specific accretion rate.  Strong radial drift for these particles is the limiting factor. The accretion efficiency of smaller particles strongly depends on the turbulent state of the disk.  Overall, these sensitivities indicate that accretion is very much dependent on the condition regarding the gaseous state (pressure gradient, turbulence) and on the material strength of the particles, which to a large extent determines the typical fragment size.

In the end, any growth scenario must be tested against observational constraints, \eg\ the size distribution of the asteroid \citep{MorbidelliEtal2009,Weidenschilling2011}, dust production rates in debris disks \citep{ShannonWu2011,KenyonBromley2008,KenyonBromley2010}, or the exoplanet census \citep{AlibertEtal2011,IdaLin2010}.  For detailed comparison, however, our toy model with its discrete, three component approximation is a rather crude tool.  But it is a great tool to conduct a preliminary exploration of the vast parameter space. This provides the user with a first, crude comparison with the observational data and a means to identify the relevant physical processes that are at work, thereby guiding further sophisticated modeling efforts.  We believe that such an interplay between a toy- and sophisticated model is a very powerful asset.

\acknowledgements{We thank Tilman Birnstiel, Kees Dullemond, Christoph Mordasini, and Satoshi Okuzumi for discussions and useful advice.  We appreciate the careful review of the referee and his/her suggestion to include \se{ext-accr}.}

\bibliography{ads,arXiv}

\appendix

\section{$P_\mathrm{col}$ in the drag-modified regime (\texttt{ND=1})}
\labelH{app:OK10}
For completeness, we provide a summary of the algorithm that \citet{OrmelKlahr2010} (OK10) provide to derive $P_\mathrm{col}$ when the equations of motion of the (small) particles are significantly affected by gas drag.  For this work we have taken the opportunity to streamline the fitting expressions for $P_\mathrm{col}$, slightly improving the correspondence to the numerical integrations.\footnote{The modifications with respect to OK10 entail the definition of the expression $\tau^\ast$ and the extension of an exponentially-declining tail to $b_\mathrm{3b}$.}

\begin{deluxetable}{lllll}
  \tablecaption{\labelH{tab:OK10}Expressions for $P_\mathrm{col}$ when \texttt{ND=1}.}
  \tablehead{
  Regime           &  Condition                                            & Impact radius\tablenotemark{a} $b_\mathrm{col}$             & Impact velocity $v_a$                     & Smoothed impact radius, $b_\mathrm{col}^\ast$
  }
  \startdata
    Settling (set)   &$\tau_\mathrm{fr}\le \tau^\ast$                        &  \Eq{bset}                                        &$3b_\mathrm{set}/2 +\zeta_w$                                               & $b_\mathrm{set} \exp\left[-(\tau_\mathrm{fr}/\tau^\ast)^{0.65}\right]$ \\
    Hyperbolic (hyp) &$\tau^\ast\le \tau_\mathrm{fr} \le \zeta_w$            &$\alpha_E \sqrt{1+6/(\alpha_E v_\mathrm{hyp}^2)}$  &$\zeta_w \sqrt{1+4\tau_\mathrm{fr}^2}\left/(1+\tau_\mathrm{fr}^2)\right.$  & $b_\mathrm{hyp}$ \\
    Three body (3b)  &$\tau_\mathrm{fr} \ge \max[\tau_\mathrm{fr}, \zeta_w]$ &$1.7\alpha_E + 1.0/\tau_\mathrm{fr}$               & 3.2                                                                       & $b_\mathrm{3b}\exp\left[-(0.7\zeta_w/\tau_\mathrm{fr})^5\right]$
  \enddata
  \tablenotetext{a}{The minimum impact radius is always the geometrical radius, $b_\mathrm{col}=\alpha_E$.}
  \tablecomments{Summary of expressions to derive $P_\mathrm{col}\equiv 2v_a b^\ast_\mathrm{col}$ when nebular drag influences the particle trajectories around the protoplanet (\texttt{ND=1}).  Expressions and parameters are normalized with Hill units: $\tau_\mathrm{fr}=T_\mathrm{fr}\Omega$, $\zeta_w=v_\mathrm{hw}/v_h$, and $\alpha_E = R_E/R_h$. The critical friction times is defined by $\tau^\ast = \min(12/\zeta_w,2)$.}
\end{deluxetable}
In the drag-modified regime, $P_\mathrm{col}$ is a function of three (dimensionless) parameters: the dimensionless friction time $\tau_\mathrm{fr} = T_\mathrm{fr}\Omega$; the `headwind parameter' $\zeta_w = v_\mathrm{hw}/v_\mathrm{h}$ (which is via $v_h$ a function of planet mass); and the planet size $\alpha_E = R_E/R_h$. The OK10 expressions cover three regimes, where the gas drag-gravity interaction is qualitatively different. From small to large $\tau_\mathrm{fr}$ these are: \sumi\ the full settling regime; \sumii\ the hyperbolic regime; \sumiii\ and the three body regime.  The regime boundaries are $\tau_\mathrm{fr}=\tau^\ast = \min(12/\zeta_w^3,2)$ and $\tau_\mathrm{fr}=\zeta_w$.  The hyperbolic regime disappears for large protoplanets.  

For each of these regimes OK10 gives expressions for the impact radius $b_\mathrm{col}$ and the approach velocity $v_a$.  For the settling regime, valid for $\tau_\mathrm{fr}\le \tau^\ast$ one obtains $b_\mathrm{col}$ by solving a cubic equation
\begin{equation}
  b^3 +\frac{2\zeta_w}{3}b^2 -8\tau_\mathrm{fr} = 0,
  \label{eq:bset}
\end{equation}
whose real solution is denoted $b_\mathrm{set}$. Similarly, we find the impact radii for the hyperbolic and three-body regime ($b_\mathrm{hyp}$ and $b_\mathrm{3b}$, respectively).  Each of these regimes, furthermore, has a different expression for the approach velocity $v_a$. These expressions are summarized in \Tb{OK10}.

The expressions for $b_\mathrm{col}$ and $v_a$ have been derived by OK10 using a simple physical model. These expressions are not continuous across boundaries.  In particular, this pertains to the behavior of $b_\mathrm{hyp}$ which is often much lower than $b_\mathrm{set}$ and $b_\mathrm{3b}$ near the boundaries.  To provide for a continuous transition one can smooth the impact radii $b_\mathrm{col}$ by appending $b_\mathrm{set}$ and $b_\mathrm{3b}$ with an exponentially-decaying tail, see the last column of \Tb{OK10}.  This causes the regimes to overlap, \eg\ the validity of the full settling regime is extended beyond $\tau_\mathrm{fr}>\tau_\mathrm{fr}^\ast$.  The smoothing exponentials are empirically determined by fitting the resulting $P_\mathrm{col}$ to the $P_\mathrm{col}$ obtained from numerical integration.

The 2D accretion rate is then given by $P_\mathrm{col} \equiv 2b_\mathrm{col}^\ast v_a$ where $b_\mathrm{col}^\ast$ is the empirical (smoothed) variant of the analytically-derived $b_\mathrm{col}$ (see \Tb{OK10}).  Thus, one obtains $P_\mathrm{set}=2b_\mathrm{set}^\ast v_\mathrm{set}$ and, similarly, $P_\mathrm{hyp}$ and $P_\mathrm{3b}$.  By virtue of the exponential continuation, multiple solutions now exist for a given $\tau_\mathrm{fr}$ and for the final $P_\mathrm{col}$ we take the maximum:
\begin{equation}
  P_\mathrm{col} = \max \left( P_\mathrm{set}, P_\mathrm{hyp}, P_\mathrm{3b} \right).
\end{equation}

\section{Calculation of the radius enhancement by atmospheres, assuming constant opacity}
\labelH{app:atmos}
We will assume that the putative atmosphere of a growing protoplanet is radiatively supported. The equations for stellar structure then read:
\begin{eqnarray}
  \labelH{eq:stel-struc0}
  P &=& \frac{k_B}{\mu m_H} \rho T \\
  \frac{dP}{dr} &=& - \frac{GM}{r^2} \rho \\
  \labelH{eq:stel-struc3}
  \frac{dT}{dr} &=& - \frac{3\kappa L_c}{64\pi \sigma_{SB}} \frac{\rho}{r^2 T^3}
\end{eqnarray}
where $P$ is the pressure, $k_B$ is Boltzmann's constant, $\mu m_H$ the molecular weight, $\rho$ the (gas) density, $T$ the temperature, $r$ the height of the atmosphere as measured from the center of the protoplanet, $\kappa$ the opacity, $L_c$ the luminosity of the protoplanet, $\sigma_\mathrm{SB}$ Stefan-Boltzmann's constant and $G$ Newton's gravitational constant.  We normalize the radial coordinate $r$ to the Bondi radius, $x=r/R_b$, where $R_b = GM/\gamma c_s^2$, $c_s^2 = P_g/\rho_g$ the isothermal sound speed of the nebula, and $\gamma=1.4$. \Eqsto{stel-struc0}{stel-struc3} are further normalized by the nebula quantities, \ie\ $p=P/P_g$, $\sigma=\rho/\rho_g$, and $\theta=T/T_g$. \Eqsto{stel-struc0}{stel-struc3}, in nondimensional form, then read
\begin{eqnarray}
  \labelH{eq:stel-struc}
  \labelH{eq:ideal}
  p &=& \sigma \theta \\
  \labelH{eq:dpdx}
  \frac{dp}{dx} &=& - \gamma \frac{\sigma}{x^2} \\
  \frac{d\theta}{dx} &=&-\gamma W_\mathrm{neb} \frac{\sigma}{\theta^3 x^2}
  \labelH{eq:dthdx}
\end{eqnarray}
where $W_\mathrm{neb}$ is the only remaining parameter:\footnote{The definition of $W_\mathrm{neb}$ differs by a factor of 4 from the $W_0$ defined by \citealt{InabaIkoma2003}.}
\begin{equation}
  W_\mathrm{neb} = \frac{3\kappa L_c}{64\pi\sigma_\mathrm{SB}} \frac{P_g}{GM T_g^4}.
\end{equation}

Dividing \eq{dthdx} by \eq{dpdx} we can readily solve for the temperature as function of the pressure
\begin{equation}
  \theta^4 = 1+4W_\mathrm{neb}(p-1),
  \labelH{eq:th4-p}
\end{equation}
where the nebular boundary condition $\theta(p=1)=1$ has been applied.  We now approximate \eq{th4-p} assuming $\theta$ is small ($\theta\ll1$; nearly isothermal) or large ($\theta\gg1$; pressure dominated).  The boundary between these regimes is set at a radius $x_1$ and a corresponding density $\sigma_1$.  By construction, both solutions join at this point.

\subsection{Limit $4W_\mathrm{neb} p \ll 1$ (nearly isothermal)}
When $4W_\mathrm{neb}p \ll 1$ \eq{th4-p} can be approximated as $\theta\approx 1+W_\mathrm{neb} (p-1)$. We then obtain
\begin{equation}
  p = \sigma\theta \approx \sigma(1+W_\mathrm{neb} (p-1)) \approx W_\mathrm{neb} \sigma^2 +\sigma(1-W_\mathrm{neb}).
  \labelH{eq:psig}
\end{equation}
With this equation and the chain rule, we write \eq{dpdx} as
\begin{equation}
  \left( 2W_\mathrm{neb} \sigma +1-W_\mathrm{neb} \right) \frac{d\sigma}{dx} = -\gamma \frac{\sigma}{x^2}.
\end{equation}
Integration gives
\begin{equation}
  2W_\mathrm{neb} \sigma + (1-W_\mathrm{neb})\log \sigma = \frac{\gamma}{x} +C_g.
\end{equation}
Here the integration constant $C_g$, obtained from $\sigma(x=1)=1$, evaluates to $C_g=2W_\mathrm{neb}-\gamma$.  We therefore find
\begin{equation}
  \frac{1}{x} \approx 1+ \frac{2W_\mathrm{neb} (\sigma-1) + \log \sigma}{\gamma}
  \labelH{eq:nearly-iso}
\end{equation}
(where we assumed $W_\mathrm{neb}\ll1$).

\subsection{Limit $4W_\mathrm{neb} p \gg 1$ (pressure dominated)}
In this case we approximate \eq{th4-p} as $\theta^4 \approx 4W_\mathrm{neb} p$.  Then, \eq{dthdx} reduces to $d\theta/dx= -\gamma W_\mathrm{neb} p /\theta^4 x^2 =-\gamma/4x^2$ and $\theta = \gamma/4x + C_1$, where $C_1$ is another integration constant.  Using \eq{ideal} we have:
\begin{equation}
  \sigma = \frac{p}{\theta} = \frac{\theta^3}{4W_\mathrm{neb}} = \frac{1}{4W_\mathrm{neb}} \left( C_1 + \frac{\gamma}{4x} \right)^3.
\end{equation}
The integration constant $C_1$ may be found from the condition $\sigma(x_1)=\sigma_1$, \ie\ $C_1+\gamma/4x_1 = (4W_\mathrm{neb}\sigma_1)^{1/3}$ and
\begin{equation}
  \sigma \approx \frac{1}{4W_\mathrm{neb}} \left( [4W_\mathrm{neb} \sigma_1]^{1/3} +\frac{\gamma}{4x} -\frac{\gamma}{4x_1} \right)^3;\quad (x \ll x_1).
  \labelH{eq:pres-sup}
\end{equation}
Thus, for $x\ll \min(1,x_1)$ the density scales as the cube of inverse radius, in agreement with previous studies \citep{Stevenson1982,InabaIkoma2003}. 

\subsection{Results}
\begin{figure}[t]
  \centering
  \includegraphics[width=8cm]{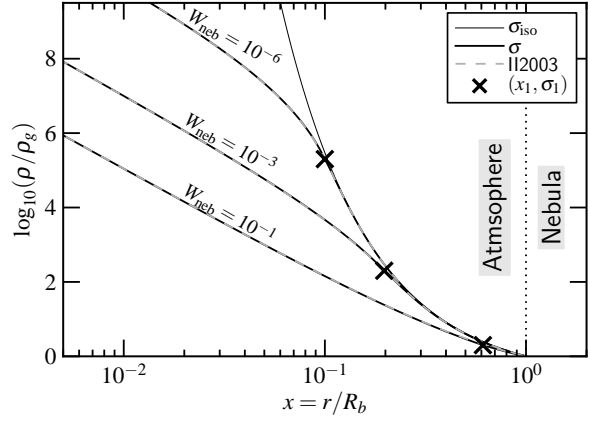}
  \caption{Solutions for the density $\sigma=\rho/\rho_g$ as function of dimensionless radius $x$ in case of constant opacity throughout the atmosphere.  At large $x$ the density follows the isothermal curve.  However, after a critical radius $x_1$ (indicated by the cross) the envelope becomes pressure-supported and the density follows a $x^{-3}$ law. Dashed-gray curves are the \citet{InabaIkoma2003} solution for the density structure.\\}
  \labelH{fig:atmos}
\end{figure}
We yet need to specify the transition between the nearly-isothermal and the pressure-dominated regimes, that is $\sigma_1$ (and corresponding $p_1$).  If we put the transition at $4W_\mathrm{neb}p_1=1$, $\sigma_1$ can be obtained from \eq{psig}: $\sigma_1 \approx p_1/(1+W_\mathrm{neb}p_1) \approx W_\mathrm{neb}^{-1}/5$.  We therefore define
\begin{equation}
  \sigma_1 \equiv \frac{1}{5W_\mathrm{neb}}
  \labelH{eq:sigma1}
\end{equation}
(from which $x_1$ follows from \eqp{nearly-iso}). The density structure calculated by our method is in excellent agreement with the full analytic solution from \citep{InabaIkoma2003}.  In \fg{atmos} we show several examples.  

Our approximation provides a 1-1 relation between the radius and the density, $\sigma=\sigma(x)$. \citet{InabaIkoma2003} calculate that a particle must experience a peak density of
\begin{equation}
  \rho_a = \frac{(6+e_h^2) s_p \rho_s}{9R_h} 
\end{equation}
in order to lose a sufficient amount of its (3-body) energy to become captured by the protoplanet.  Inverting the expressions for $\sigma(x)$ then gives us a direct solution for the capture radius ($x_a$) in closed form.  That is, using \eq{nearly-iso} for densities $\sigma_a=\rho_a/\rho_g<\sigma_1$ and
\begin{equation}
  \frac{1}{x_a} = \frac{1}{x_1} +\frac{4}{\gamma}(4W_\mathrm{neb})^{1/3} \left( \sigma_a^{1/3} -\sigma_1^{1/3} \right)
\end{equation}
for densities $\sigma_a$ larger than $\sigma_1$.  

\begin{figure}[t]
  \centering
  \includegraphics[width=8cm]{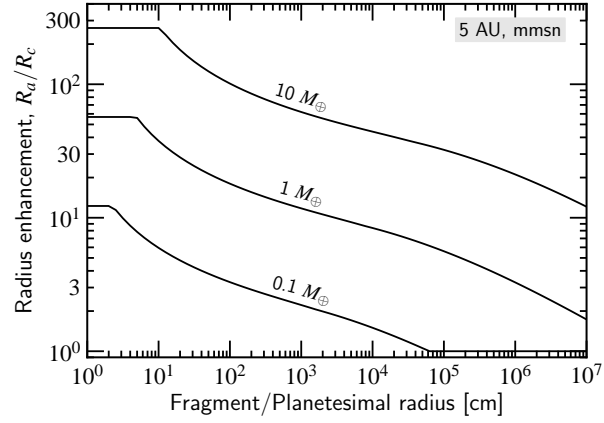}
  \caption{Solutions for the capture radius $R_a$, normalized to the core radius, as function of particle size for several protoplanet masses. The protoplanet is placed at a disk radius of 5 AU and accretes particles at a rate of $1\ M_\oplus\ \mathrm{Myr}^{-1}$. The eccentricity of the particles is fixed at $e_h=4$, independent of their size.}
  \labelH{fig:atmos2}
\end{figure}
In \fg{atmos2} we plot the radius enhancement factor of the embryo ($R_a/R_c$) as function of particle size protoplanets of mass 0.1, 1.0, and 10 $M_\oplus$.  For simplicity, we have fixed the Hill eccentricity at $4e_h$, although in reality this will be a function of particle radius too.  Likewise, the accretion rate is fixed at $\dot M = 1\ M_\oplus\ \mathrm{yr}^{-1}$.  With these parameters the $W_\mathrm{neb}$ values are $2.8\times10^{-4}$, $1.3\times10^{-4}$, and $5.9\times10^{-5}$, respectively. The enhancement or $R_a$ is largest for smaller particles as they are most affected by the drag.  The increase in $R_a$ continues until the `boundary' of the atmosphere is hit, which is given by the Bondi radius.  The radius increase is also a steep function of protoplanet mass.  \Fg{atmos2} can be compared to Fig.\ 2 of \citet{KobayashiEtal2011}. 

\section{Stirring by turbulent density fluctuations}
\labelH{app:densflucstir}
Planets (and planetesimals) are scattered by gas (over)densities induced by turbulence.  The statistical distribution of these torque fluctuations is determined by two key quantities: the amplitude of the rms-fluctuations, $\sigma_\Gamma$ and the correlation time $\tau_c$.  These allow us to define a diffusion coefficient, $D_j=\sigma_\Gamma^2 \tau_c$ such that the induced eccentricity change becomes
\begin{equation}
  e \simeq \frac{\Delta j}{j} = \frac{\sqrt{D_j t}}{j}
\end{equation}
where $j=a^2 \Omega$ is the specific angular momentum.  Squaring and differentiating with respect to time gives the stirring rate
\begin{equation}
  \frac{de^2}{dt} = \frac{D_j}{j^2} = \left( \frac{\sigma_\Gamma}{j} \right)^2 \tau_c.
\end{equation}
The magnitude of the fluctuations, $\sigma_\Gamma$, is redefined in terms of a nondimensionless parameter $\gamma_t$:
\begin{equation}
  \sigma_\Gamma = \frac{\tilde{C} \Sigma \gamma_t a^4 \Omega^2}{M_\star},
\end{equation}
\citep{BaruteauLin2010} with $\tilde{C}=2.4\times10^2$. If we assume that $\tau_c=\Omega^{-1}$ we find
\begin{equation}
  \label{eq:de2dt-ts}
  \frac{de^2}{dt} = \left(\frac{\tilde{C}\gamma_t a^2\Sigma_g}{M_\star} \right)^2 \Omega.
\end{equation}
\citet{BaruteauLin2010} relate $\gamma_t$ to the diffusion parameter $\alpha_\mathrm{ss}$, $\gamma_t=8.5\times10^{-2} \alpha_\mathrm{ss}^{1/2} H_g/a$. Inserting this expression in \eq{de2dt-ts}, these numerical constants we obtain a stirring rate of
\begin{equation}
  \frac{de^2}{dt} \approx 4\times10^{2} \alpha_\mathrm{ss} \left( \frac{H_g a \Sigma_g}{M_\star}\right)^2 \Omega
\end{equation}
and a turbulent stirring timescale of:
\begin{equation}
  T_\mathrm{ts} 
  = \frac{2e^2}{de^2/dt} 
  \approx e^2\frac{5\times10^{-3}}{\alpha_\mathrm{ss}} \left( \frac{H_g a \Sigma_g}{M_\star}\right)^{-2} \Omega^{-1}.
\end{equation}
It is instructive to compare the turbulent and viscous stirring timescales (\eqp{Tvs}):
\begin{eqnarray}
  \frac{T_\mathrm{ts}}{T_\mathrm{vs}} 
  &=&\ \frac{5\times10^{-3}}{4\pi\tilde{b}\alpha_\mathrm{ss}} \frac{e^2 R_h}{e_h^2 a_0} P_\mathrm{vs} \left( \frac{H_g a_0 \Sigma_g}{M_\star}\right)^{-2}
   \approx   \frac{10^{-5}}{\alpha_\mathrm{ss}} \frac{M_E}{M_\star} P_\mathrm{vs} \left( \frac{H_g a_0 \Sigma_g}{M_\star}\right)^{-2}\\
   &\approx &\ 1.0 \left( \frac{\alpha_\mathrm{ss}}{10^{-4}} \right)^{-1} \frac{P_\mathrm{vs}}{10} \left( \frac{M_E/M_\star}{10^{-8}} \right) \left( \frac{H_g/a}{0.1} \right)^{-2} \left( \frac{\Sigma_g a^2/M_\star}{10^{-3}} \right)^{-2},
\end{eqnarray}
where we used that $e/e_h=R_h/a_0 = (M_E/3M_\star)^3$. Thus, turbulent stirring may over dominate viscous stirring for small protoplanets (small $M_E$), massive disks (large $\Sigma_g$) at large disk radii (where the flaring is stronger), and large $\alpha_\mathrm{ss}$.

\end{document}